# A Navier-Stokes asymptotic preserving Direct Simulation Monte Carlo method for multi-species gas flows

Fei Fei[*]

*School of Aerospace Engineering, Huazhong University of Science and Technology, 430074 Wuhan, China*
*Corresponding author: ffei@hust.edu.cn

**Abstract:** This paper introduces a new asymptotic-preserving Monte Carlo (APMC) method for simulating multi-species gas flows. This method decomposes the collision operator of the traditional APMC methods into macro and micro collision parts: the macro collision part corresponds to the first-order Chapman-Enskog (CE) expansion, solved with a second-order scheme; In contrast, the micro collision part represents the reminder of high-order nonequilibrium terms, calculated following the traditional APMC methods. As a result, this new APMC method preserves Navier-Stokes asymptotics and second-order accuracy in the fluid limit, while maintaining the advantage of the traditional APMC methods in the kinetic limit. Therefore, it is suitable for multi-species gas flows that span multiple scales due to superior resolution and accuracy in both continuum and rarefied flow regimes. Moreover, directly constructed based on the DSMC collision solver, this new APMC method circumvents the complexities of modeling and computing kinetic models. This feature makes it easily applicable to multi-species gas flows with intricate collision models and more than two species. Finally, a coupling scheme with DSMC has been developed to extend the applicability of this new APMC method to multi-species gas flows with disparate masses.

## 1. Introduction

Multi-species gas flows wildly exist in aerospace engineering and micro-electro-mechanical systems (MEMS). Typically, in these applications, multi-species gas flows range from the continuum to the rarefied flow regimes, therefore, an appropriate numerical scheme should be capable of tackling this multi-scale challenge efficiently. The direct simulation Monte Carlo (DSMC) method [1] has been well established for the gas flows in the rarefied flow regime. However, as the gas flow approaches the continuum flow regime, employing DSMC to achieve high-resolution results becomes exceedingly expansive due to stringent restrictions on cell size and time step. An alternative approach for computing multi-scale gas flows is to couple the DSMC and CFD methods through domain decomposition, known as the hybrid CFD-DSMC method [2, 3]. The advantage of the hybrid method is its capacity to balance accuracy and efficiency across different scales. Nevertheless, since the DSMC and CFD methods are individually derived from microscopic and macroscopic perspectives, the exchange of information at their interface [4] and the determination of consistent transport properties [5] are still challenging problems.

From a computational standpoint, the efficiency of DSMC in the continuum flow regime is hindered by the substantial computational expense associated with binary collisions. To address this, kinetic approaches, such as the Bhatnagar-Gross-Krook (BGK) [6] and Fokker-Planck (FP) [7] models, have been introduced to simplify the Boltzmann collision term. The BGK model replaces the Boltzmann collision term with relaxation processes, making it more tractable for analysis and computation. Currently, the applicability of the BGK model has extended beyond monoatomic gases to include polyatomic [8, 9], multi-species [10-19], and chemical reactive [20, 21] gas flows. Considering multi-species gas flows, existing BGK models can be categorized into single-relaxation [12-14] and multi-relaxation [15-19] types. While the single-relaxation model is less complicated to implement, the multi-relaxation model offers a more accurate depiction of the energy and momentum exchange rates. Based on these BGK models, several deterministic or stochastic numerical methods have been proposed for simulating multi-species gas flows. These methods, such as the unified gas kinetic scheme (UGKS) [22, 23], discrete unified gas kinetic scheme (DUGKS) [24], general synthetic iterative scheme (GSIS) [25], and stochastic particle methods [26, 27], are not limited by the mean free path and the collision time, thereby providing higher efficiency for multi-scale flows compared to DSMC. In addition to the BGK model, the Fokker-Planck model is another kinetic approach, which treats the sequence of molecular collisions as a continuous stochastic process. Numerical schemes based on the FP model [28, 29] can also overcome the limitations imposed by collision scales. Like BGK models, the FP model has also been extended to multi-species gas flows [30-34]. However, most BGK and FP models, as simplifications of detailed molecular collisions, cannot reproduce all transport coefficients, such as thermal diffusion. Although the multi-relaxation approach can theoretically describe interactions between species, determining adjustable parameters to make the transport processes consistent with those obtained from DSMC collision models or Navier-Stokes (CFD) solvers remains challenging, particularly for gas mixtures with more than two species [19]. Moreover, BGK and FP models are questionable for highly nonequilibrium gas flows [35, 36]. This defect also limits their applicability to multi-scale gas flows that span various flow regimes.

Recently, asymptotic-preserving (AP) schemes have been developed to circumvent the above difficulties in kinetic models. These schemes directly tackle the Boltzmann collision term, including the time-relaxed [37], exponential Runge–Kutta [38], and asymptotic-preserving Monte Carlo [39] methods. They have also been extended to handle the multi-species Boltzmann equation [40, 41]. The fundamental concept of these AP schemes is to decompose the Boltzmann integral into a stiff linear part and a less stiff nonlinear part [38]. Typically, this stiff linear part is represented by a BGK-type relaxation operator [42], ensuring the asymptotic preserving of the Euler equations.

However, this AP property cannot accurately reproduce the transport processes such as mixture viscosity, conductivity, and diffusion in the fluid limit, which are crucial for applications involving multi-species gas flows. Moreover, the AP property of the Euler equations can lead to reduced accuracy in the fluid limit [43], introducing significant numerical dissipation. To overcome these limitations, a Navier-Stokes asymptotic-preserving DSMC method has recently been developed based on the time-relaxed Monte Carlo (TRMC) scheme [44]. By decomposing the TRMC collision operator into micro and macro collision parts, and employing appropriate time discretization for the latter, this new TRMC method satisfies the AP property of the Navier-Stokes equations and second-order accuracy in the fluid limit. Furthermore, during the construction of this scheme, no kinetic approximations for the DSMC collision model are required, thus making it potentially extendable to more complex gas simulations, such as those involving multi-species gas flows.

This paper aims to extend the previously proposed Navier-Stokes asymptotic-preserving DSMC method for the multi-species gases while recovering all transport coefficients and maintaining second-order accuracy in the fluid limit. The remainder of the paper is organized as follows: Sections 2 and 3 first provide brief reviews of the multi-species Boltzmann equation and existing APMC schemes. In Section 4.1, a new APMC method for multi-species gases is developed based on the TRMC method. The properties of this new method are discussed in Section 4.2. Furthermore, to address multi-species gas flows with disparate masses, a novel coupling method that combines the newly proposed APMC scheme with DSMC is introduced in Section 4.3. In Section 5, several numerical test cases are computed to validate the method's high resolution and efficiency for multi-scale flows. Finally, the conclusion is given in Section 6.

## 2. The multi-species Boltzmann equation

### 2.1 The Boltzmann equation

The Boltzmann equation for multi-species gases is expressed as

$$\frac{\partial f_s}{\partial t} + \boldsymbol{V} \cdot \frac{\partial f_s}{\partial \boldsymbol{x}} = \frac{1}{\varepsilon} Q_s(f, f), \qquad (2.1)$$

where $f_s(\boldsymbol{V}; \boldsymbol{x}, t)$ represents the probability density function (PDF) of the s-th species with velocity $\boldsymbol{V} \in \mathbb{R}^3$ at position $\boldsymbol{x}$ and time $t$. For the sake of simplicity, this work focuses on the monoatomic gas. Consequently, the collision term of the Boltzmann equation is written as

$$Q_s(f,f) = \sum_{k=1}^{N_s} Q_{sk}(f,f), \qquad (2.2a)$$

$$Q_{sk}(f,f) = \int_{\mathbb{R}^3} \int_{\mathbb{S}^2} [f_s(\boldsymbol{V}')f_k(\boldsymbol{V}_1') - f_s(\boldsymbol{V})f_k(\boldsymbol{V}_1)] g\sigma_{sk} d\Omega d\boldsymbol{V}_1, \qquad (2.2b)$$

where $\boldsymbol{V}$ and $\boldsymbol{V}_1$ are the pre-collisional velocities of the collision pair, and $\boldsymbol{V}'$ and $\boldsymbol{V}_1'$ are their post-collisional velocities. The relative velocity of the collided molecules is denoted by $g = |\boldsymbol{V} - \boldsymbol{V}_1|$. $\sigma_{sk}$ represents the differential cross-section of the binary collision between the s-th and k-th species. $\Omega$ denotes the solid angle of the sphere $S^2$. $N_s$ denotes the total number of species. Generally, the Boltzmann collision term can be further split into two parts

$$Q_{sk}(f,f) = Q_{sk}^+ - f_s Q_{sk}^-, \tag{2.3}$$

where

$$\begin{cases} Q_{sk}^+ = \int_{\mathbb{R}^3} \int_{\mathbb{S}^2} f_s(\boldsymbol{V}') f_k(\boldsymbol{V}_1') g\sigma_{sk} d\Omega d\boldsymbol{V}_1 \\ Q_{sk}^- = \int_{\mathbb{R}^3} \int_{\mathbb{S}^2} f_k(\boldsymbol{V}_1) g\sigma_{sk} d\Omega d\boldsymbol{V}_1 \end{cases}, \tag{2.4}$$

which are defined as the gain and loss terms of the Boltzmann collision integral, respectively.

Note that the Boltzmann equation (2.1) is presented using dimensionless quantities, which are introduced as follows:

$$\hat{x} = \frac{x}{x_*}, \hat{V} = \frac{V}{\bar{C}}, \hat{t} = \frac{t}{x_*/\bar{C}}, \hat{g} = \frac{g}{\sqrt{2}\bar{C}}, \hat{\sigma} = \frac{\sigma}{\pi d^2}, \hat{f} = \frac{f}{n_*/\bar{C}^3}, \tag{2.5}$$

where $x_*$ and $n_*$ are the characteristic length and number density, respectively. $\bar{C} = \sqrt{kT_*/m_*}$ is defined as the characteristic velocity, where $T_*$ and $m_*$ denote the characteristic temperature and mass. $d$ is the reference molecular diameter. For simplicity, the notation for dimensionless variables, i.e., an upper triangle, will be omitted in the subsequent discussion. Using these dimensionless quantities, we can derive the Knudsen number as $\varepsilon = 1/(x_* \sqrt{2} \pi d^2 n_*)$.

In addition, the macroscopic quantities for species s, such as the number density $n_s$, mass density $\rho_s$, average velocity $\boldsymbol{u}_s$, temperature $T_s$, shear stress $\Gamma_{s,ij}$ and heat flux $q_{s,i}$, are calculated by integrating the PDF of the s-th species over the phase space using the following expressions:

$$\begin{cases} n_s = \int f_s d\boldsymbol{V}; \rho_s = m_s n_s; \\ \boldsymbol{u}_s = \frac{1}{n_s} \int \boldsymbol{V} f_s d\boldsymbol{V}; T_s = \frac{m_s}{3n_s} \int C^2 f_s d\boldsymbol{V}; \\ \Gamma_{s,ij} = m_s \int C_{<i} C_{j>} f_s d\boldsymbol{V}; q_{s,i} = \frac{1}{2} m_s \int C_i C^2 f_s d\boldsymbol{V} \end{cases}, \tag{2.6}$$

where $\boldsymbol{C} = \boldsymbol{V} - \boldsymbol{u}$ denotes the peculiar velocity of molecules, and $\boldsymbol{u}$ is the mean velocity of the mixture, given by $\boldsymbol{u} = \sum_{k=1}^{N_s} \rho_k \boldsymbol{u}_k / \rho$. The total mass density is defined as $\rho = \sum_{k=1}^{N_s} \rho_k$. $C_{<i} C_{j>}$ denotes the symmetric and trace-free part of the tensor $C_i C_j$. Moreover, the other mean variables, such as the mean temperature $T$, total number density $n$, total pressure $p$, total shear stress $\Gamma_{ij}$ and total heat flux $q_i$, are defined as,

$$\begin{cases} T = \frac{1}{n} \sum_{k=1}^{N_s} n_k T_k; n = \sum_{k=1}^{N_s} n_k; p = nT; \\ \Gamma_{ij} = \sum_{k=1}^{N_s} \Gamma_{k,ij}; q_i = \sum_{k=1}^{N_s} q_{k,i} \end{cases}. \tag{2.7}$$

The Boltzmann equation adheres to the conservation law and H-theory; therefore, for a homogenous system, the PDF of the mixture gas will ultimately relax to a Maxwellian distribution, given by

$$M_{s,eq}(\boldsymbol{V}; \boldsymbol{x}, t) = n_s \left(\frac{m_s}{2\pi T}\right)^{3/2} exp\left(-\frac{m_s C^2}{2T}\right). \tag{2.8}$$

Taking moments of Eq. (2.1) with this Maxwellian distribution and summarizing all the species yields the Euler equations, i.e.,

$$\begin{cases} \frac{D\rho}{Dt} + \rho \frac{\partial u_i}{\partial x_i} = 0 \\ \frac{Du_i}{Dt} + \frac{1}{\rho}\frac{\partial p}{\partial x_i} = 0 \\ \frac{DT}{Dt} + \frac{p}{c_v \rho}\frac{\partial u_i}{\partial x_i} = 0 \end{cases}, \quad (2.9)$$

where $D/Dt = \partial/\partial t + u_i \partial/\partial x_i$ is the convective time derivative, and $c_v$ denotes the specific heat at constant volume for a monoatomic gas.

## 2.2 The DSMC method

Due to the high dimensionality of the Boltzmann equation, the direct simulation Monte Carlo (DSMC) method is popularly used to simulate rarefied gas flows involving multi-species. In DSMC, the Boltzmann equation is split into a convection part and a collision part, i.e.,

**Convection**: $\frac{\partial f_s}{\partial t} + \mathbf{V} \cdot \frac{\partial f_s}{\partial \mathbf{x}} = 0,$ (2.10a)

**Collision**: $\frac{\partial f_s}{\partial t} = \frac{1}{\varepsilon} Q_s(f, f).$ (2.10b)

The solution of the convection part can be written as

$$f_s^*(\mathbf{V}; \mathbf{x}) = \mathcal{T}_{\Delta t}(f_s^n) = f_s^n(\mathbf{V}; \mathbf{x} - \mathbf{V}\Delta t), \quad (2.11)$$

where $f_s^n$ and $f_s^*$ denote the PDFs before and after the convection step, respectively. $\mathcal{T}_{\Delta t}(f)$ is defined as the convection operator with time step $\Delta t$. In practice, the PDF of molecular velocities is represented by a set of simulated particles in DSMC, i.e.,

$$f_s(\mathbf{V}; \mathbf{x}) = \frac{W_P}{\mathbb{V}} \sum_{\alpha=1}^{N_{c,s}} \delta(\mathbf{V} - \mathbf{M}^{(\alpha)}) \delta(\mathbf{x} - \mathbf{X}^{(\alpha)}), \quad (2.12)$$

where $\mathbf{M}^{(\alpha)}$ and $\mathbf{X}^{(\alpha)}$ represent the velocity and position of the simulated particle with index of $\alpha$. The volume of the system is denoted by $\mathbb{V}$. Consequently, the number of simulated particles of s-th species is obtained as $N_{c,s} = n_s \mathbb{V}/W_P$, where $W_P$ is the number of real molecules represented by a simulated particle. In DSMC, the solution of the convection part (2.11) is implemented by directly tracing the simulated particles, i.e., keeping the velocity unchanged and updating the position as $\mathbf{X}^{(\alpha),*} = \mathbf{X}^{(\alpha),n} + \mathbf{M}^{(\alpha),n}\Delta t$.

The DSMC method calculates binary collisions after the particle convection step. For the Nanbu-Babovsky scheme [45], the governing equation of the collision part (2.10b) can be rewritten as

$\frac{\partial f_s}{\partial t} = \frac{1}{\varepsilon}[P_s(f, f) - \beta f_s]$, where (2.13)

$P_s(f, f) = Q_s(f, f) + \beta f_s.$ (2.14)

Solving Eq. (2.13) with a forward Euler scheme, one can obtain the collision operator of DSMC $\mathcal{C}_{\Delta t}^{DSMC}(f)$ as

$$f_s^{n+1}(\mathbf{V}; \mathbf{x}) = \mathcal{C}_{\Delta t}^{DSMC}(f_s^*) = \left(1 - \frac{\beta \Delta t}{\varepsilon}\right) f_s^* + \frac{\beta \Delta t}{\varepsilon} \frac{P_s(f^*, f^*)}{\beta}. \quad (2.15)$$

Equation (2.15) indicates that the collision process, in the DSMC simulation, divides the simulated particles into two groups: a collisionless group with a proportion of $(1 - \beta \Delta t / \varepsilon)$ and a collisional group with a proportion of $\beta \Delta t / \varepsilon$. The collided particle pairs are selected from the second group to satisfy the actual mean collision rate, and updated velocities according to the dynamics of binary collisions. Besides the Nanbu-Babovsky scheme, the DSMC collision operator can also be implemented using Bird's no-time counter (NTC) scheme, and detailed implementations of NTC can be found in reference [1].

To ensure the positivity of each term in Eq. (2.15), the time step of DSMC is restricted by the mean collision time, i.e., $\Delta t < \varepsilon/\beta$; at the same time, the parameter $\beta$ should be carefully chosen to preserve the positivity of each term in $P_s(f,f)$. Similar to ref. [40], $\beta$ is set as a constant here, which satisfies

$$\beta = constant > sup_{V,s}|Q_s^-|. \tag{2.16}$$

In particle methods, $sup_{V,s}|Q_s^-|$ can be estimated by $n(g\sigma_{T,sk})_{max}$, which is calculated as the maximum value among all collided particle pairs. Here, $\sigma_{T,sk}$ refers to the total collision cross-section for the s-th and k-th species, defined as $\sigma_{T,sk} = \int_{\mathbb{S}^2} \sigma_{sk} d\Omega$.

As the Knudsen number decreases, i.e., $\varepsilon/\beta \to 0$, the time step of DSMC becomes much smaller than the characteristic time for macro flows. This results in a significant decrease in the computational efficiency. To overcome this restriction, several asymptotic preserving Monte Carlo (APMC) schemes have been proposed, such as the Exponential Runge-Kutta (ExpRK), asymptotic-preserving Monte Carlo (AP-DSMC), time-relaxed Monte Carlo (TRMC) method. A brief review of these APMC schemes will be provided in the following.

## 3. Review of the asymptotic preserving Monte Carlo method for multi-species gas flows

### 3.1 The Exponential Runge-Kutta Method

Li and Yang first extended the ExpRK method for simulating multi-species gas flows [40]. Although the ExpRK method is presented as a deterministic scheme in ref. [40], it can also be implemented using stochastic particles. In the ExpRK method, a splitting scheme similar to Eq. (2.10) is utilized, and the convection part is solved in the same manner as in DSMC.

To avoid stiffness, the ExpRK method rewrites the collision part as,

$$\frac{\partial[(f_s - M_{s,eq})e^{\beta t/\varepsilon}]}{\partial t} = \frac{1}{\varepsilon}[P_s(f,f) - \beta M_{s,eq}]e^{\beta t/\varepsilon}, \tag{3.1}$$

which is then solved by a Runge-Kutta scheme. For the lowest order, the numerical solution of Eq. (3.1) is given by

$$f_s^{n+1}(\mathbf{V}; \mathbf{x}) = C_{\Delta t}^{ExpRK}(f_s^*) = A_0 f_s^* + A_1 \frac{P_s(f^*,f^*)}{\beta} + A_2 M_{s,eq}^*, \tag{3.2}$$

where $A_0 = e^{-\frac{\beta \Delta t}{\varepsilon}}$, $A_1 = \frac{\beta \Delta t}{\varepsilon} e^{-\frac{\beta \Delta t}{\varepsilon}}$ and $A_2 = 1 - A_0 - A_1$. $C_{\Delta t}^{ExpRK}(f)$ denotes the collision operator of the ExpRK method.

## 3.2 The AP-DSMC method

The asymptotic-preserving Monte Carlo method was proposed by Ren et.al. [39], and then extended to multi-species gas flows by Zhang, et.al. [41]. In the AP-DSMC method, the splitting scheme as shown in Eq. (2.10) is also employed. However, the AP-DSMC scheme reconstructs the collision part using the successive BGK penalty method. Therefore, Eq. (2.10b) is reconstructed as

$$\frac{\partial f_s}{\partial t} = \frac{1}{\varepsilon}[Q_s(f,f) - \beta(M_{s,eq} - f_s)] + \frac{\beta}{\varepsilon}(M_{s,eq} - f_s). \tag{3.3}$$

The first term on the right-hand side of Eq. (3.3) is less stiff and solved explicitly. The second term represents the linear stiff BGK penalty term. In the AP-DSMC method, the BGK penalty term is divided into $L$ subintervals and solved implicitly within each subinterval. For instance, when $L=2$, the numerical solution of the collision part for the AP-DSMC scheme is given by [41],

$$f_s^{n+1}(\boldsymbol{V};\boldsymbol{x}) = C_{\Delta t}^{AP-DSMC}(f_s^*) = A_0 f_s^* + A_1 \frac{P_s(f^*,f^*)}{\beta} + A_2 M_{s,eq}^*. \tag{3.4}$$

where $A_0 = \left(1 + \frac{\beta \Delta t}{2\varepsilon}\right)^{-2}$, $A_1 = \frac{\beta \Delta t}{\varepsilon}\left(1 + \frac{\beta \Delta t}{2\varepsilon}\right)^{-2}$ and $A_2 = 1 - A_0 - A_1$.

## 3.3 The time-relaxed Monte Carlo method

Besides the ExpRK and AP-DSMC methods, another notable asymptotic-preserving Monte Carlo method is the time-relaxed Monte Carlo method, which was proposed by Pareschi and Russo [37]. Although the TRMC method only considers single-species gas flows originally, extending it to the simulations of multi-species gas flows is straightforward. The convection part of the TRMC scheme is the same as that of the ExpRK and AP-DSMC methods. In contrast, the TRMC method transforms the collision part as

$$\frac{\partial F_s}{\partial t'} = \frac{1}{\beta} P_s(F,F), \tag{3.5}$$

where $t' = 1 - e^{-\beta t/\varepsilon}$, $F_s(\boldsymbol{V};\boldsymbol{x},t') = f_s(\boldsymbol{V};\boldsymbol{x},t)e^{\beta t/\varepsilon}$, and the bilinearity of $P_s(f,f)$ is employed. The numerical solution of Eq. (3.5) can be derived from the Wild Sum together with a Maxwellian truncation. The first-order collision operator of the TRMC method is given by

$$f_s^{n+1}(\boldsymbol{V};\boldsymbol{x}) = C_{\Delta t}^{TRMC}(f_s^*) = A_0 f_s^* + A_1 \frac{P_s(f^*,f^*)}{\beta} + A_2 M_{s,eq}^*, \tag{3.6}$$

where $A_0 = e^{-\beta \Delta t/\varepsilon}$, $A_1 = 1 - A_0 - A_2$ and $A_2 = (1 - e^{-\beta \Delta t/\varepsilon})^3$.

Although different in theory, the collision operators of the APMC methods share the same form when considering only the lowest order scheme, as illustrated in Eqs. (3.2), (3.4) and (3.6). Consequently, their implementations are also similar. In the computation of the collision part, simulated particles are divided into three groups according to the coefficients $A_0$, $A_1$ and $A_2$. The velocities of particles in the first group remain unchanged, while those in the second group are updated through the DSMC collision process. Lastly, the velocities of the remaining particles are resampled from the Maxwellian distribution. Since $A_2 \to 1$ as $\varepsilon \to 0$ for these APMC methods, their PDF solutions converge to the Maxwellian distribution $M^*_{s,eq}$. This ensures the AP property of the Euler equations.

## 4. The direct simulation Monte Carlo method preserving the Navier-Stokes asymptotics

As discussed in Section 3, traditional asymptotic preserving Monte Carlo methods for multi-species gas flows are limited to the AP property of the Euler equations and, therefore, can not correctly reproduce transport processes in the fluid limit. Moreover, due to the Euler AP property, the accuracy of the traditional APMC methods will reduce to first-order in the fluid limit [43, 44]. In this section, a novel APMC method is introduced to address these issues. To summarize, the main features of this proposed APMC method would be:

- Asymptotic preserving property of the NS equations;
- Maintaining second-order accuracy in the fluid limit and the same accuracy as the traditional APMC methods in the kinetic limit;
- Flexible in employing various DSMC collision models.

In addition, although this newly proposed APMC method is constructed based on the TRMC method in this work, the underlying strategy can be readily applied to other APMC methods, such as the ExpRK and AP-DSMC schemes. This adaptability is due to the similar structure of the collision operator among these methods. Moreover, for simplicity, this proposed scheme is named the asymptotic accurate and preserving Monte Carlo method, i.e., the AAPMC method, in the following.

### 4.1 The numerical scheme of the AAPMC method

4.1.1. The basic scheme

In the AAPMC method, we also split the Boltzmann equation as shown in Eq. (2.10). Then, using the Strang splitting, the general form of the AAPMC scheme is given by

$$f_s^{n+1} = \mathcal{C}^{AAPMC}_{\Delta t/2}\{\mathcal{T}_{\Delta t}[\mathcal{C}^{AAPMC}_{\Delta t/2}(f_s^n)]\}, \quad (4.1)$$

where $\mathcal{C}^{AAPMC}_{\Delta t}(f)$ denotes its collision operator. These collision operators can be combined between sequential time steps, then we have

$$\begin{aligned} f_s^{n+1} &= \mathcal{C}^{AAPMC}_{\Delta t/2} \mathcal{T}_{\Delta t} \mathcal{C}^{AAPMC}_{\Delta t/2} \mathcal{C}^{AAPMC}_{\Delta t/2} \mathcal{T}_{\Delta t} \cdots \mathcal{T}_{\Delta t} \mathcal{C}^{AAPMC}_{\Delta t/2} \mathcal{C}^{AAPMC}_{\Delta t/2} \mathcal{T}_{\Delta t} \mathcal{C}^{AAPMC}_{\Delta t/2}(f_s^0) \\ &= \mathcal{C}^{AAPMC}_{\Delta t/2} \mathcal{T}_{\Delta t} \mathcal{C}^{AAPMC}_{\Delta t} \mathcal{T}_{\Delta t} \cdots \mathcal{T}_{\Delta t} \mathcal{C}^{AAPMC}_{\Delta t} \mathcal{T}_{\Delta t} \mathcal{C}^{AAPMC}_{\Delta t/2}(f_s^0). \end{aligned} \quad (4.2)$$

In practice, the convection and collision steps of the AAPMC scheme can be solved sequentially with an entire time step the same as those in DSMC.

In the AAPMC method, the numerical solution for the convection step is consistent with those in other APMC schemes, as provided by Eq. (2.11).

For the collision step, a micro-macro decomposition is employed to preserve the Navier-Stokes asymptotics in the fluid limit, similar to the previous AAP-TRMC method [44]. Consequently, the collision operator of the AAPMC scheme is presented as

$$\mathcal{C}^{AAPMC}_{\Delta t}(f_s) = \underbrace{\mathcal{C}^{APMC}_{\Delta t}(f_s) - \mathcal{C}^{APMC,(1)}_{\Delta t}(f_s)}_{micro} + \underbrace{\mathcal{C}^{fluid,(1)}_{\Delta t}(f_s)}_{macro}, \qquad (4.3)$$

where $\mathcal{C}^{APMC,(1)}_{\Delta t}(f)$ and $\mathcal{C}^{fluid,(1)}_{\Delta t}(f)$ denote the first-order Chapman-Enskog (CE) expansion of a traditional APMC collision operator and a designed macro collision operator, respectively. As well known, the CE expansion expresses the PDF of molecular velocities as a series corresponding to the Knudsen number, i.e.,

$$f_s = f_s^{(0)} + f_s^{(1)} + f_s^{(2)} + \cdots, \qquad (4.4)$$

where $f_s^{(0)} = M_{s,eq}$, and $f_s^{(1)}$ and $f_s^{(2)}$ denote the first and second-order of the CE expansions, respectively.

In Equation (4.3), $\mathcal{C}^{APMC}_{\Delta t}(f)$ can be any collision operators of existing APMC methods. Without loss of generality, the TRMC collision operator is employed in the present work. As shown in Eq. (4.3), the AAPMC method first divides the collision operator of the traditional APMC methods into micro and macro collision parts. For the macro collision part, a designed macro collision operator is used to replace the first-order CE expansion of the traditional APMC collision operator. Additionally, the micro collision part, referring to the first two terms on the right-hand side of Eq. (4.3), represents the remaining high-order terms of the CE expansion.

The detailed construction for the micro and macro collision parts within the AAPMC scheme is described as follows:

First, we estimate the solution after the convection step. According to the last term of Eq. (4.2), given $f_s^n$ as the initial condition, we assume the PDF before the convection step as

$$\mathcal{C}^{AAPMC}_{\Delta t/2}(f_s^n) = f_s^n + \frac{\Delta t}{2\varepsilon} Q_s(f^n, f^n). \qquad (4.5)$$

Then, using the convection step solution given by Eq. (2.11), the PDF solution after the particle convection step is obtained as

$$f_s^* = \mathcal{T}_{\Delta t} \mathcal{C}^{AAPMC}_{\Delta t/2}(f_s^n) = f_s^{n+1} - \frac{\Delta t}{2\varepsilon} Q_s(f^{n+1}, f^{n+1}) + O(\Delta t^3), \qquad (4.6)$$

where the Taylor expansion and the Boltzmann equation are utilized in the derivation.

Second, the macro collision part $\mathcal{C}^{fluid}_{\Delta t}(f)$ is constructed using a second-order scheme, which is written as

$$\mathcal{C}^{fluid}_{\Delta t}(f_s) = f_s + \frac{\Delta t}{\varepsilon} Q_s(f, f) + \frac{\Delta t^2}{2\varepsilon} \frac{\partial Q_s(f, f)}{\partial t}. \qquad (4.7)$$

Substituting the solution of the convection step obtained by Eq. (4.6), we obtain the numerical solution of the macro collision part as

$$\mathcal{C}^{fluid}_{\Delta t}(f_s^*) = f_s^{n+1} + \frac{\Delta t}{2\varepsilon} Q_s(f^{n+1}, f^{n+1}) + O(\Delta t^3). \qquad (4.8)$$

The first-order CE expansion for Eq. (4.8) is written as

$$C_{\Delta t}^{fluid,(1)}(f_s^*) = f_s^{n+1,(1)} + \frac{\Delta t}{2\varepsilon} Q_s^{(1)}(f^{n+1}) + O(\Delta t^3), \tag{4.9}$$

where $Q_s^{(1)}(f^{n+1}) = Q_s(f^{n+1,(1)}, M_{eq}^{n+1}) + Q_s(M_{eq}^{n+1}, f^{n+1,(1)})$ denotes the first-order Chapman-Enskog expansion for the Boltzmann collision term.

Third, using the collision operator of the traditional TRMC method presented in Eq. (3.6) and the solution of the convection step, the micro collision part of the AAPMC method is organized as

$$C_{\Delta t}^{APMC}(f_s^*) - C_{\Delta t}^{APMC,(1)}(f_s^*) = \begin{cases} A_0 f_s^* + A_1 \frac{P_s(f^*,f^*)}{\beta} + A_2 M_{s,eq}^* \\ -\left[(1-A_2)f_s^{*,(1)} + A_1 \frac{Q_s^{(1)}(f^*)}{\beta}\right] \end{cases}, \tag{4.10}$$

where $M_{s,eq}^*$ refers to the Maxwellian distribution after the convection step. Owing to the conservation of the Boltzmann collision integral, it follows that $M_{s,eq}^* = M_{s,eq}^{n+1}$.

Finally, substituting Eqs. (4.9) and (4.10) into Eq. (4.3), the collision operator of the AAPMC method is obtained as

$$C_{\Delta t}^{AAPMC}(f_s^*) = A_0 f_s^* + A_1 \frac{P_s(f^*,f^*)}{\beta} + A_2 (M_{s,eq}^{n+1} + f_s^\alpha), \tag{4.11}$$

where

$$f_s^\alpha = \begin{cases} f_s^{n+1,(1)} + \left(1 - \frac{A_2}{2} - A_1 \frac{\varepsilon}{\beta \Delta t}\right) \frac{1}{A_2} \frac{\Delta t}{\varepsilon} Q_s^{(1)}(f^{n+1}) \\ + \frac{A_1}{A_2} \frac{\Delta t}{2\varepsilon} \frac{1}{\beta} Q_s^{(1)} \left[Q_s^{(1)}(f^{n+1})\right] \end{cases}. \tag{4.12}$$

The collision operator of the AAPMC method is similar to that of the traditional APMC schemes, except $M_{s,eq}^*$ is replaced by $(M_{s,eq}^* + f_s^\alpha)$. Therefore, in practice, we only need to change the sampling process for particles belonging to the third group. To accurately sample particle velocities satisfied $(M_{s,eq}^{n+1} + f_s^\alpha)$, the associated PDFs, i.e., $f_s^{n+1,(1)}$, $\frac{1}{\varepsilon} Q_s^{(1)}(f^{n+1})$ and $\frac{1}{\varepsilon\beta} Q_s^{(1)}\left[Q_s^{(1)}(f^{n+1})\right]$, should be first specialized, which will be introduced in the subsequent sections. Furthermore, while the collision operator presented in Eq. (4.11) is based on the first-order TRMC method, it is also feasible to extend this approach to the second-order TRMC method (see Appendix A).

4.1.2. The estimation of the PDF for the first-order CE expansion

The first-order CE expansion can be derived from the Chapman-Enskog theory. However, due to the complexity of its expression, computing the first-order CE expansion for systems involving more than two species is challenging, especially when DSMC collision models are considered. To address this issue, we introduce a local DSMC collision solver, which directly computes the first-order CE expansion for mixture gases. The details of this approach are presented as follows:

First, the first-order Chapman-Enskog expansion $f_s^{n+1,(1)}$ can be written as [46]

$$f_s^{n+1,(1)} = M_{s,eq}^{n+1} \left\{ \frac{D_{s,i}^{(1)} C_i}{\rho_s \theta_s} + \frac{1}{2} \frac{\Gamma_{s,ij}^{(1)} C_{<i} C_{j>}}{\rho_s \theta_s^2} + \frac{2}{5\rho_s \theta_s^2} \left[q_{s,i}^{(1)} - \frac{5}{2}\theta_s D_{s,i}^{(1)}\right] C_i \left(\frac{C^2}{2\theta_s} - \frac{5}{2}\right) \right\}, \tag{4.13}$$

where $\theta_s$ and $D_{s,i}$ are defined as $\theta_s = T/m_s$ and $D_{s,i} = \rho_s(u_{s,i} - u_i)$, respectively. The terms $D_{s,i}^{(1)}$, $\Gamma_{s,ij}^{(1)}$ and $q_{s,i}^{(1)}$ represent the modified diffusion velocity, shear stress and heat flux, respectively, corresponding to the first-order CE expansion.

Eq. (4.13) demonstrates that the first-order CE expansion is a deviational distribution function from the Maxwellian. This deviational distribution function can be represented by a set of signed simulated particles as

$$f_s^{n+1,(1)}(\mathbf{V}) = \frac{1}{\mathbb{V}} \sum_{\alpha=1}^{N_{dev,s}} sgn\left[f_s^{n+1,(1)}(\mathbf{V})\right] \delta(\mathbf{V} - \mathbf{M}^{(\alpha)}), \tag{4.14}$$

where $sgn[f_s^{(1)}(\mathbf{V})]$ denotes the sign of the particle weight, which is either 1 or -1. The total number of signed simulated particles is given by $N_{dev,s} = \frac{\mathbb{V}}{W_P} \int \left|f_s^{n+1,(1)}(\mathbf{V})\right| d\mathbf{V}$. The particle velocity $\mathbf{M}^{(\alpha)}$ is sampled from the distribution (4.13) using the acceptance-rejection algorithm. There are

$$N_{dev,s}^{(P)} = \sum_{\alpha=1}^{N_{dev,s}} max\left\{sgn\left[f_s^{n+1,(1)}(\mathbf{M}^{(\alpha)})\right], 0\right\}, \tag{4.15a}$$

positive particles and

$$N_{dev,s}^{(N)} = \sum_{\alpha=1}^{N_{dev,s}} max\left\{-sgn\left[f_s^{n+1,(1)}(\mathbf{M}^{(\alpha)})\right], 0\right\}, \tag{4.15b}$$

negative particles. Since $\int f_s^{n+1,(1)}(\mathbf{V})d\mathbf{V} = 0$, it is noted that $N_{dev,s}^{(P)} = N_{dev,s}^{(N)}$ on average.

Second, according to the Chapman-Enskog theory, the first-order CE expansion $f_s^{n+1,(1)}$ is governed by

$$\frac{\partial M_{s,eq}^{n+1}}{\partial t} + V_i \frac{\partial M_{s,eq}^{n+1}}{\partial x_i} = \frac{1}{\varepsilon}\left[Q_s\left(f^{n+1,(1)}, M_{eq}^{n+1}\right) + Q_s\left(M_{eq}^{n+1}, f^{n+1,(1)}\right)\right]. \tag{4.16}$$

Giving a Maxwellian distribution, $f_s^{n+1,(1)}$ can be solved from Eq. (4.16) by using an iteration approach, i.e.,

$$f_{s,m}^{n+1,(1)} = C_{\Delta t^*}^{dev\_DSMC}\left(f_{s,m-1}^{n+1,(1)}\right) - \Delta t^* \left(\frac{\partial M_{s,eq}^{n+1}}{\partial t} + V_i \frac{\partial M_{s,eq}^{n+1}}{\partial x_i}\right), \tag{4.17}$$

where the index $m$ denotes the iteration step, and the initial value is set as $f_{s,0}^{n+1,(1)}$. $\Delta t^*$ denotes the time step for iterations, here it is selected as $\varepsilon/\beta$. In Eq. (4.17), $C_{\Delta t^*}^{dev\_DSMC}$ denotes a DSMC collision operator to compute the linear Boltzmann collision integral on the right-hand side of Eq. (4.16), which is constructed as

$$C_{\Delta t^*}^{dev\_DSMC}\left(f_{s,m-1}^{n+1,(1)}\right) = f_{s,m-1}^{n+1,(1)} + \frac{\Delta t^*}{\varepsilon}\left[Q_s\left(f_{m-1}^{n+1,(1)}, M_{eq}^{n+1}\right) + Q_s\left(M_{eq}^{n+1}, f_{m-1}^{n+1,(1)}\right)\right]. \tag{4.18}$$

As the first-order CE expansion $f_s^{n+1,(1)}$ is represented by signed simulated particles that sampled from a deviational distribution function, Eq. (4.18) should be solved using a deviational DSMC algorithm [47-49] instead of the standard DSMC method. The detailed implementation of the deviational DSMC algorithm for solving the linear Boltzmann collision integral is presented in Appendix B.

The implementations of the local DSMC collision solver, which is used to estimate the first-order CE expansion $f_s^{n+1,(1)}$, are summarized in Algorithm 4.1. Additionally, $M$ represents the number of iterations. For steady flows, $M$ can be set to one for simplicity.

**Algorithm 4.1** The local DSMC collision solver to estimate the first-order CE expansion.

---

For the computational cell $j$ and time step $n$:

1: Assign and sample signed simulated particles according to the first-order CE expansion $f_{s,0}^{n+1,(1)}$ as given by (4.13), with initial variables of $\boldsymbol{D}_{s,0}^{(1)}$, $\boldsymbol{\Gamma}_{s,0}^{(1)}$ and $\boldsymbol{q}_{s,0}^{(1)}$ obtained from the previous time step and initial Maxwellian distribution $M_{s,eq}^{n+1}$ computed based on mean variables after the convection step.

2: **for** m=1 to M **do**

3:    Implement the deviational DSMC algorithm to obtain $C_{\Delta t^*}^{dev\_DSMC}\left(f_{s,m-1}^{n+1,(1)}\right)$ (see Appendix B);

4: Sample and assign additional signed simulated particles satisfied the deviational PDF $-\Delta t^*\left(\frac{\partial M_{s,eq}^{n+1}}{\partial t} + V_i \frac{\partial M_{s,eq}^{n+1}}{\partial x_i}\right)$, then the PDF solution for the next iteration $f_{s,m}^{n+1,(1)}$ is obtained using Eq. (4.17).

6: **end for**

7: Average the signed simulated particles to obtain $\boldsymbol{D}_{s,M}^{(1)}$, $\boldsymbol{\Gamma}_{s,M}^{(1)}$ and $\boldsymbol{q}_{s,M}^{(1)}$, which are used to compute the first-order CE expansion $f_s^{n+1,(1)}$ according to Eq. (4.13).

---

4.1.3. The estimation of the PDF for $\frac{1}{\varepsilon}Q_s^{(1)}(f^{n+1})$ and $\frac{1}{\varepsilon\beta}Q_s^{(1)}\left[Q_s^{(1)}(f^{n+1})\right]$

As shown in Eq. (4.12), the PDF $f_s^\alpha$ is composed of three parts. The first part is the first-order CE expansion $f_s^{n+1,(1)}$, presented by Eq. (4.13), which can be computed using the local DSMC collision solver described in Section 4.1.2. The second part is the first-order CE expansion of the collision term $\frac{1}{\varepsilon}Q_s^{(1)}(f^{n+1})$, given by

$$\frac{1}{\varepsilon}Q_s^{(1)}(f^{n+1}) = \frac{\partial M_{s,eq}^{n+1}}{\partial t} + V_i \frac{\partial M_{s,eq}^{n+1}}{\partial x_i} = M_{s,eq}^{n+1}\left\{\begin{array}{l}\frac{1}{\theta_s}\left[\boldsymbol{C}\cdot(\boldsymbol{C}\cdot\nabla)\boldsymbol{u}-\frac{1}{3}C^2\nabla\cdot\boldsymbol{u}\right]+\boldsymbol{C}\cdot\nabla T\frac{1}{T}\left(\frac{C^2}{2\theta_s}-\frac{5}{2}\right)\\ +\frac{C}{\theta_s}\left[\frac{1}{\rho_s}\nabla(n_s T)-\frac{1}{\rho}\nabla p\right]\end{array}\right\}. \quad (4.19)$$

The third part remains unknown at the current stage. Since this part also satisfies the first-order CE expansion and the conservative property, we assume that it can be approximated by the same expression as Eq. (4.12), however, with specialized variables $\boldsymbol{D}_{s,ex}^{(1)}$, $\boldsymbol{\Gamma}_{s,ex}^{(1)}$ and $\boldsymbol{q}_{s,ex}^{(1)}$. For this purpose, we introduce a deviational DSMC collision process to determine the last term in $f_s^\alpha$, i.e., $\frac{1}{\varepsilon\beta}Q_s^{(1)}\left[Q_s^{(1)}(f^{n+1})\right]$ in Algorithm 4.2.

**Algorithm 4.2** Deviational DSMC collision process to estimate $\frac{1}{\varepsilon\beta}Q_s^{(1)}\left[Q_s^{(1)}(f^{n+1})\right]$

---

For the computational cell $j$ and time step $n$:

1: Initialize signed simulated particles based on the deviational PDF $\frac{1}{\varepsilon}Q_s^{(1)}(f^{n+1})$;

2: Implement the deviational DSMC algorithm (see Appendix B) for a time step $\frac{\varepsilon}{\beta}$, obtain $C_{\varepsilon/\beta}^{dev\_DSMC}\left[\frac{1}{\varepsilon}Q_s^{(1)}(f^{n+1})\right] = \frac{1}{\varepsilon}Q_s^{(1)}(f^{n+1}) + \frac{1}{\varepsilon\beta}Q_s^{(1)}\left[Q_s^{(1)}(f^{n+1})\right]$;

3: Sample and assign additional signed simulated particles satisfied the PDF $-\frac{1}{\varepsilon}Q_s^{(1)}(f^{n+1})$, then the targeted PDF $f_{s,ex}^{(1)} = C_{\varepsilon/\beta}^{dev\_DSMC}\left(\frac{1}{\varepsilon}Q_s^{(1)}\right) - \frac{1}{\varepsilon}Q_s^{(1)}$ is obtined.

4: Average the PDF obtained in step 3 to compute corresponding variables, such as $\boldsymbol{D}_{s,ex}^{(1)}$, $\boldsymbol{\Gamma}_{s,ex}^{(1)}$ and $\boldsymbol{q}_{s,ex}^{(1)}$;

5: Finally, the PDF $\frac{1}{\varepsilon\beta}Q_s^{(1)}\left[Q_s^{(1)}(f^{n+1})\right]$ is determined according to Eq. (4.12).

### 4.1.4. The implementation of the AAPMC method for multi-species gas flows

We summarize the implementation of the AAPMC method for multi-species gas flows in Algorithm 4.3 and depict the schematic representation of this scheme in Figure 1. Although the implementation described here is based on the first-order AAPMC method, the analogous extension to the second-order AAPMC method, presented in Appendix A, can be easily derived following Algorithm 4.3.

**Algorithm 4.3** The implementation of the AAPMC method.

For the computational cell $j$:

1: Initialize simulated particles according to the initial PDF $f_s^0$.

2: **for** n=1 to $N_{final}$ **do**

3:   **Particle convection**.

   3.1 Update positions of simulated particles based on the solution of the convection step, as Eq. (2.11). The treatment of boundary conditions follows the same procedure as in DSMC. The PDF after the convection step is denoted as $f_s^*$;

   3.2 Compute the ensemble average of $f_s^*$ to obtain the mean variables, which are used to determine the Maxwellian distribution $M_{s,eq}^{n+1}$;

4:   **Local DSMC collision solver**.

   4.2 Estimate the first-order CE expansion of the collision term $\frac{1}{\varepsilon}Q_s^{(1)}(f^{n+1})$ based on the expression of Eq. (4.19);

   4.2 Estimate the first-order CE expansion $f_s^{n+1,(1)}$ using Algorithm 4.1;

   4.2 Estimate the PDF $\frac{1}{\varepsilon\beta}Q_s^{(1)}\left[Q_s^{(1)}(f^{n+1})\right]$ using Algorithm 4.2.

5:   **AAPMC collision solver**.

   5.1 Determine the parameter $\beta$ using Eq. (2.16);

   5.2 Compute the coefficients $A_0$, $A_1$ and $A_2$, then randomly divide simulated particles into three groups proportional to these coefficients;

   5.3 Particles with the ratio $A_0$ maintain their velocities unchanged;

   5.4 Particles with the ratio $A_1$ undergo the DSMC collision operator $P_s(f^*,f^*)/\beta$;

   5.5 Particles with the ratio $A_2$ resample their velocities according to distribution $(M_{s,eq}^{n+1} + f_s^\alpha)$, as shown in Eq. (4.12).

6: **Sampling and averaging**.
   Compute the ensemble average of the simulated particles before and after the collision step to obtain the macroscopic quantities.
7: **end for**
8: Output results.

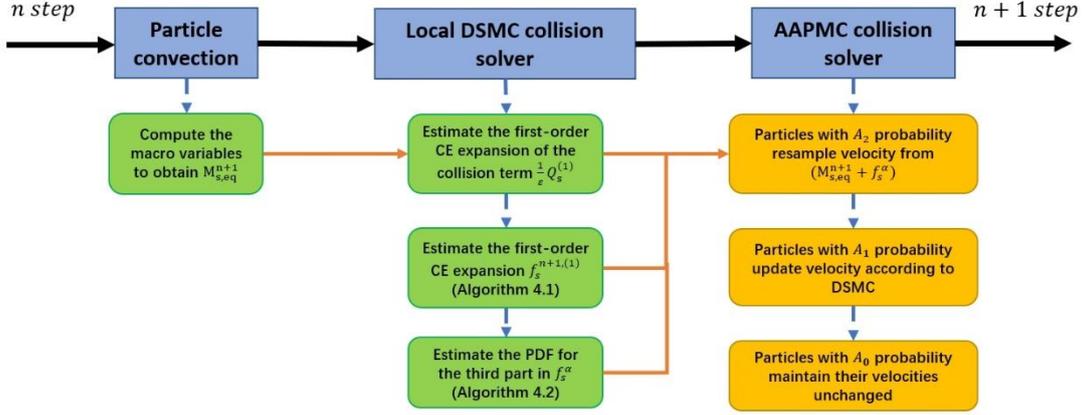

Figure 1. The schematic diagram of the AAPMC scheme.

**4.2 The numerical properties of the AAPMC method**

4.2.1. Positivity

The AAPMC method is composed of the particle convection and collision steps. During the convection step, the positivity of the distribution is always maintained. For the collision step, as described by Eq. (4.11), the first two terms, which are analogous to those in other APMC schemes, obviously preserve positivity. However, in the last term, the AAPMC method replaces the Maxwellian distribution $M_{s,eq}^*$ by a modified PDF $(M_{s,eq}^{n+1} + f_s^\alpha)$, which may include negative parts. We estimate its positivity in the fluid and kinetic limits in the following.

In the fluid limit, as $\beta \Delta t / \varepsilon \gg 1$, in Eq. (4.12), the coefficients $A_0$ and $A_1$ approach 0, while $A_2$ approaches 1. Consequently, $(M_{s,eq}^{n+1} + f_s^\alpha)$ can be approximated as

$$M_{s,eq}^{n+1} + f_s^\alpha \approx M_{s,eq}^{n+1} + f_s^{n+1,(1)} + \frac{\Delta t}{2\varepsilon} Q_s^{(1)}(f^{n+1}). \tag{4.20}$$

Furthermore, using Eqs. (4.13) and (4.19), Eq. (4.20) can be rewritten as

$$M_{s,eq}^{n+1} + f_s^{n+1,(1)} + \frac{\Delta t}{2\varepsilon} Q_s^{(1)}(f^{n+1}) \approx M_{s,eq}^{n+1} \left\{ \begin{array}{l} 1 + C_i \left( \tau_s + \frac{\Delta t}{2} \right) \frac{1}{n_s} \frac{\partial n_s}{\partial x_i} \\ + \frac{C_{<i} C_{j>}}{\theta_s} \left( \tau_s + \frac{\Delta t}{2} \right) \frac{\partial u_i}{\partial x_j} \\ + C_i \left( \frac{C^2}{2\theta_s} - \frac{5}{2} \right) \left( \tau_s + \frac{\Delta t}{2} \right) \frac{1}{T} \frac{\partial T}{\partial x_i} \end{array} \right\}. \tag{4.21}$$

Here, according to the kinetic theory, we approximate the diffusion, viscosity and thermal conductivity coefficients for species $s$ as $\tau_s p_s / \rho_s$, $\tau_s p_s$ and $c_p \tau_s p_s$, respectively, where $\tau_s$ denotes the mean collision time and $c_p$ is specific heat at constant pressure. The positivity of the distribution given by Eq. (4.21) can be estimated following Ref. [52]. Similarly, a nonequilibrium parameter B is defined as

$$B = \left(\tau_s + \frac{\Delta t}{2}\right) max\left(\left|\sqrt{\frac{2T}{m_s}}\frac{\nabla n_s}{n_s}\right|, |\nabla u|, \left|\frac{5}{2}\sqrt{\frac{2T}{m_s}}\frac{\nabla T}{T}\right|\right). \tag{4.22}$$

The PDF, as shown in Eq. (4.21), will maintain positivity, if this parameter is bounded and small enough. Specifically, if $B < 0.1$, this PDF is negative about 0.05% of the time [52]. In the fluid limit where $\beta \Delta t / \varepsilon \gg 1$, the upper bound of the nonequilibrium parameter can be evaluated as

$$B \sim \left(\tau_s + \frac{\Delta t}{2}\right)\frac{c}{\Delta x}\frac{\Delta x}{x_*} < \left(\tau_s + \frac{\Delta t}{2}\right)\frac{|u|+c}{\Delta x}\frac{\Delta x}{x_*} \approx CFL \frac{\Delta x}{x_*}, \tag{4.23}$$

where $c$ represents the sound speed, $\Delta x$ denotes the cell size and $CFL$ is the CFL number. Generally, the CFL condition requires that $CFL < 1$, on the other hand, refined resolutions of fluid-structure demand that $\Delta x / x_*$ should be much smaller than the unit. Therefore, the nonequilibrium parameter is small enough to ensure the positivity of PDF given by Eq. (4.21) in most situations.

In the kinetic limit, as $\beta \Delta t / \varepsilon \ll 1$, the coefficient $A_2$ approaches 0, which is a higher-order term of $\beta \Delta t / \varepsilon$ compared to $A_0$ and $A_1$. As a result, though the modified PDF $(M_{s,eq}^{n+1} + f_s^\alpha)$ may not guarantee positivity, its influence for the entire system is negligible. In practice, during sampling, the AAPMC method will exclude particle velocities resulting in a negative distribution.

4.2.2. Accuracy

The numerical solutions of the AAPMC method at time step $n+1$ can be computed as the average of the PDFs obtained before and after the particle collision step, given by Eqs. (4.6) and (4.11), respectively. In the fluid limit, as $\beta \Delta t / \varepsilon \gg 1$, the PDF after the particle collision is approximated by Eq. (4.20). Consequently, the PDF solution at time step $n+1$ can be estimated as

$$f_s^{n+1} = M_{s,eq}^{n+1} + f_s^{n+1,(1)} + O(\Delta t^3), \tag{4.24}$$

This indicates that the AAPMC method achieves second-order accuracy in time within the fluid limit.

In the kinetic limit, as $\beta \Delta t / \varepsilon \ll 1$ and $A_2 \to 0$, the AAPMC method becomes identical to the traditional APMC methods, therefore, they achieves the same level of accuracy.

4.2.3. Asymptotic preserving property

Equation (4.24) also demonstrates that the proposed AAPMC method satisfies the asymptotic-preserving property of the NS equations in the fluid limit.

**4.3 The extension of the AAPMC method for disparate masses**

Disparate masses introduce additional time scales into multi-species gas flows. For instance, in a binary mixture gas, species 1 and 2 represent the light and heavy species, respectively, with $m_1 < m_2$. The relaxation times for the shear stress and heat flux of the heavy species are $\sqrt{m_2/m_1}$ times larger than those of the light species. Furthermore, the temperature difference between the species decays $\sqrt{m_2/m_1}$ times more slowly than the shear stress and heat flux of the heavy species [51]. Generally, a higher degree of nonequilibrium is closely related to a larger relaxation time for high-order moments, such as the shear stress, heat flux, and temperature difference. Consequently, as the mass ratio increases, the heavier species deviate further from the equilibrium, particularly in the temperature difference, which is more significant in multi-species gas flows with larger disparate masses.

4.3.1. The AAPMC-DSMC coupling scheme

As discussed in the previous section, the proposed AAPMC scheme satisfies the asymptotic-preserving property of the NS equations. Specifically, for a fixed time step $\Delta t$, if the Knudsen number $\varepsilon/\beta \to 0$, the PDF solution of the AAPMC collision operator approaches its fluid limit, i.e., $f_s^{n+1} \to C_{\Delta t}^{fluid,(1)}(f_s^*)$; in contrast, if $\varepsilon/\beta \to \infty$, the AAPMC collision operator reduces to the traditional APMC schemes, approaching the DSMC solver in the kinetic limit, i.e., $f_s^{n+1} \to C_{\Delta t}^{DSMC}(f_s^*)$. Since $\beta$ is chosen as the supremum of $|Q_s^-|$ as shown in Eq. (2.16), we can estimate the time scale of $(\varepsilon/\beta)$ as

$$\frac{\varepsilon}{\beta} = \frac{\varepsilon}{n(g\sigma_{T,sk})_{max}} \approx \frac{\varepsilon}{n(\sigma_{T,sk})_{max}}\sqrt{\frac{m_{min}}{T}} \sim \tau_{min}. \qquad (4.25)$$

Here, $\tau_{min}$ denotes the mean collision time of the lightest species, normalized by the characteristic time of the fluid field, i.e., $x_*/\bar{C}$, where $m_{min}$ denotes the mass of the lightest species. We note that the time scale of $(\varepsilon/\beta)$ is identical to the relaxation time of the shear stress and heat flux for the lightest species [46]. For comparable masses, the relaxation times of all species are similar, thus the time scale of the lightest species $\tau_{min}$ is sufficient to represent the non-equilibrium of the entire system. However, additional time scales emerge due to disparate masses, and the standard AAPMC scheme developed in section 4.1 may struggle to properly handle all the scales simultaneously for large disparate masses. Taking Figure 2(a) as an example, we consider a gas mixture with $N_s$ species. Their relaxation times can be ordered as $\tau_1 < \cdots < \tau_{k-1} < \tau_k < \cdots < \tau_{N_s}$. By introducing a nonequilibrium criterion $\tau_{cr}$, we can separate $N_s$ species into two groups: the first group belongs to the continuum flow regime ($1 \leq s \leq k$) and the second group belongs to the rarefied flow regime ($k + 1 < s \leq N_s$), where $k \in [1, N_s]$. Assuming $\tau_{cr} > \tau_1$, the numerical solutions of the standard AAPMC collision operator for all species would tend to the fluid limit, even the heavy species within $k + 1 < s \leq N_s$ may be far from equilibrium. Therefore, the standard AAPMC method may underestimate the rarefication for the heavy species and becomes inaccurate for multi-species gas flows with large disparate masses.

To address the challenge caused by the multiple time scales with disparate masses, a coupling method combined with DSMC is proposed to extend the standard AAPMC scheme. As shown in Fig. (2b), the basic idea is to divide the gas species into continuum and rarefied groups using a nonequilibrium criterion. For species within the continuum group, the standard AAPMC scheme is implemented as usual; conversely, for species within the rarefied group, the DSMC collision operator is employed to capture nonequilibrium phenomena, particularly for the heavy species. The schematic diagram of this coupling scheme is presented in Fig. 3.

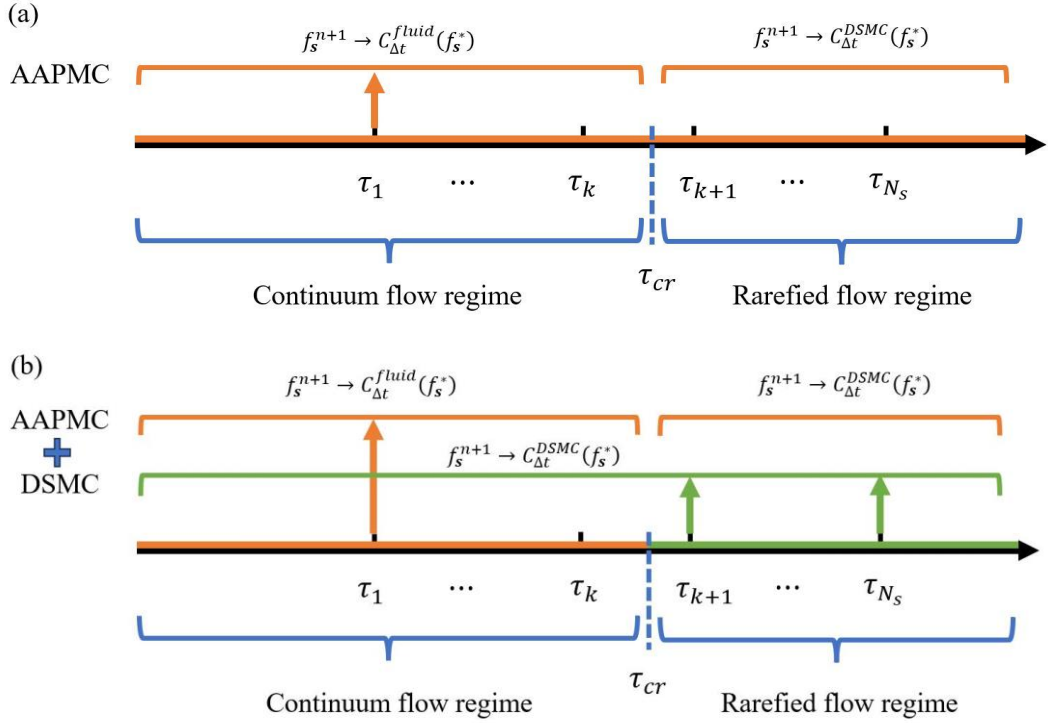

Figure 2. The time scale analysis of the (a) standard AAPMC and (b) AAPMC-DSMC coupling schemes for multi-species gas flows with disparate masses.

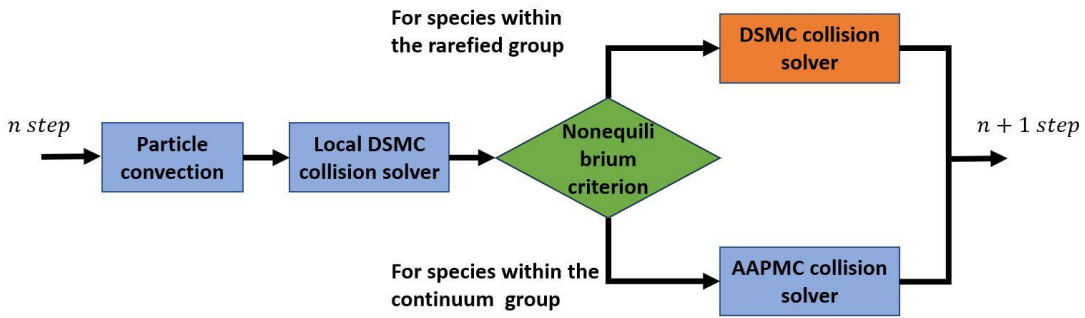

Figure 3. The schematic diagram of the AAPMC-DSMC coupling scheme for disparate masses.

4.3.2. The nonequilibrium criterion

This section introduces a nonequilibrium criterion for the AAPMC-DSMC coupling method. Eq. (4.13) indicates that high-order moments, such as diffusion velocity, shear stress, and heat flux, determine the first-order CE expansion. However, as the magnitude of these high-order moments increases, the validation of the first-order CE expansion would be doubtful. For example, distributions could become negative, which implies significant departures from the equilibrium state. Therefore, the dimensionless forms of the shear stress and heat flux have been used to assess the degree of nonequilibrium, as in ref. [53]. Here, we utilize a similar nonequilibrium parameter $B_s$ for the s-th species, i.e.,

$$B_s = max\left(\left|\sqrt{\frac{2T}{m_s}}\frac{D_{s,i}^{(1)}}{p_s}\right|, \left|\frac{\Gamma_{s,ij}^{(1)}}{p_s}\right|, \left|\sqrt{\frac{2m_s}{T}}\frac{q_{s,i}^{(1)}}{p_s}\right|\right). \tag{4.26}$$

This nonequilibrium parameter is also associated with the Knudsen number. Similar to Eq. (4.21), we approximate the first-order CE expansion of diffusion velocity, shear stress, and heat flux using gradients of mean variables, then estimating the nonequilibrium parameter by

$$B_s = max\left(\left|\frac{\rho_s}{p_s}\sqrt{\frac{T}{m_s}}\frac{\tau_s p_s}{\rho_s}\frac{1}{n_s}\frac{\partial n_s}{\partial x_i}\right|, \left|\frac{1}{p_s}\tau_s p_s\frac{\partial u_i}{\partial x_j}\right|, \left|\frac{1}{p_s}\sqrt{\frac{T}{m_s}}\tau_s p_s\frac{1}{T}\frac{\partial T}{\partial x_i}\right|\right)$$
$$\sim max\left(\left|\lambda_s\frac{1}{n_s}\frac{\partial n_s}{\partial x_i}\right|, \left|Ma_s\lambda_s\frac{1}{u_i}\frac{\partial u_i}{\partial x_j}\right|, \left|\lambda_s\frac{1}{T}\frac{\partial T}{\partial x_i}\right|\right) = max(Kn_s^n, Ma_s Kn_s^u, Kn_s^T), \tag{4.27}$$

where $Kn_s^n$, $Kn_s^u$ and $Kn_s^T$ denote the Knudsen number defined in terms of the characteristic lengths of number density, mean velocity and temperature, respectively. $\lambda_s$ denotes the mean free path of species $s$, and $Ma_s = u/\sqrt{T/m_s}$. It is noted that the nonequilibrium parameter $B_s$ is analogous to the gradient-length local (GLL) Knudsen number suggested by Wang and Boyd [54]. Moreover, to ensure the positivity of the first-order CE expansion, it is required that $B_s < 0.1$. Considering this requirement, we set the nonequilibrium criterion to $B_s = 0.1$ in this coupling scheme.

4.3.3. The implementation of the AAPMC-DSMC coupling method for disparate masses

The detailed implementation of the AAPMC-DSMC coupling scheme is summarized in Algorithm 4.4.

**Algorithm 4.4** The implementation of the AAPMC-DSMC coupling scheme.

| For the computational cell $j$: |
|---|
| 1: Assign and sample simulated particles according to the initial PDF $f_s^0$. |
| 2: **for** n=1 to $N_{final}$ **do** |
| 3:    **Particle convection**. The same as step 3 of Algorithm 4.3. |
| 4:    **Local DSMC collision solver**. The same as step 4 of Algorithm 4.3. |
| 5:    **Nonequilibrium criterion**. Calculate nonequilibrium parameter $B_s$ using Eq. (4.26). |
| 5:    **Particle collision:** |
|      5.1 If $B_s < 0.1$ for the s-th species, the AAPMC collision solver is |

implemented as step 5 of Algorithm 4.3.

    5.2 If $B_s \geq 0.1$ for the s-th species, the DSMC collision solver is implemented.

6: **Sampling and averaging**. The same as step 6 of Algorithm 4.3.

8: **end for**

9: Output results.

---

### 4.3.4 Conservation

In the AAPMC-DSMC coupling scheme, gas species are categorized into continuum and rarefied groups based on the degree of non-equilibrium. Although both the standard AAPMC and DSMC collision solvers conserve mass, momentum, and energy, the exchange rates of the momentum and energy between continuum and rarefied groups are different for these two collision operators. This discrepancy arises because the standard AAPMC collision solver is governed by the smallest relaxation time, this may lead to an underestimation of nonequilibrium for heavier species, as discussed in section 4.3.1. Therefore, to guarantee the momentum and energy conservation for the AAPMC-DSMC coupling scheme, we employ a linear transformation [55, 56] for the velocities of simulated particles belonging to the continuum group after the collision step.

## 5. Numerical results

The AAPMC method and its coupling scheme with DSMC are validated through several benchmark cases in this section. Exponentially weighted moving time averaging [7] was employed to reduce the statistical noise for the ensemble average of macroscopic variables. Furthermore, a second-order interpolation is utilized in the collision step to obtain the macro variables at particle positions. As a result, both the temporal and spatial accuracy can achieve second-order in the fluid limit for the AAPMC method. In the following, the mean free path for the multi-species gas is defined as

$$\lambda = \sum_{k=1}^{N_s} \lambda_k \chi_k \text{ and } \lambda_k = \left[\sum_{i=1}^{N_s} n_i \sigma_{ki} \sqrt{1 + \frac{m_k}{m_i}}\right]^{-1}, \quad (5.1)$$

where $\chi_k = n_k/n$ is the species concentration. As the Knudsen number and mass ratio decrease, the AAPMC-DSMC coupling method will reduce to the standard AAPMC method. Therefore, we don't distinguish the standard AAPMC and its coupling method in this section. Moreover, the AAPMC1 and AAPMC2 schemes refer to the first and second-order AAPMC methods. In the AAPMC1 scheme, the first-order TRMC collision operator (TRMC1) is employed. The AAPMC2 scheme is defined similarly.

### 5.1 Mass diffusion

The one-dimensional mass diffusion of an argon-helium mixture is first calculated within the computational domain $[0, L]$. Two reservoirs of the helium-argon mixture are established at the inner ($x = 0$) and outer ($x = L$) boundaries, with the species concentrations set to $\chi_{He}(0) = 0.7$ and $\chi_{He}(L) = 0.3$, respectively. The temperature and number density of the mixture gas in both reservoirs are maintained $300\,K$ and $1.0 \times 10^{21}\,m^{-3}$. Three Knudsen numbers, such as 0.5, 0.05, and 0.005, are simulated using DSMC, TRMC, and the proposed AAPMC methods. Additionally, the initial mean free path of the mixture gas is $2.23 \times 10^{-3}\,m$. For the DSMC method, according to the restriction on cell size, refined grids of 50, 200 and 2000 uniform cells are employed for three Kn numbers, respectively. However, much fewer cells are used in the TRMC and AAPMC methods, i.e., 50 uniform grids for all cases. Moreover, the CFL condition is employed, and the CFL number is set to 0.3.

Figure 4 compares the species concentration and diffusion velocity profiles. The variable hard sphere (VHS) model is utilized. It is observed that, due to the asymptotic preserving property of the NS equations, the AAPMC method exhibits higher accuracy than the traditional TRMC method at Kn=0.005 and 0.05. As the Knudsen number increases to 0.5, the results of the AAPMC and TRMC methods converge with those obtained from DSMC. Additionally, the second-order AAPMC and TRMC methods are more accurate than the first-order ones in the kinetic limit [44]. Therefore, for Kn=0.05 and 0.5, we have employed the AAPMC2 and TRMC2 schemes. Detailed comparisons between the first and second-order schemes are further demonstrated through simulations of Fourier and Couette flows.

Figure 5 presents a comparison of species concentration and diffusion velocity profiles using the variable soft sphere (VSS) model. We observe that the AAPMC method yields identical results to those of DSMC and exhibits higher accuracy than the traditional TRMC method, which is consistent with the findings from the VHS model. This case implies that the flexible utilization of DSMC collision models can significantly broaden the applicability of the AAPMC method to complex gas flows.

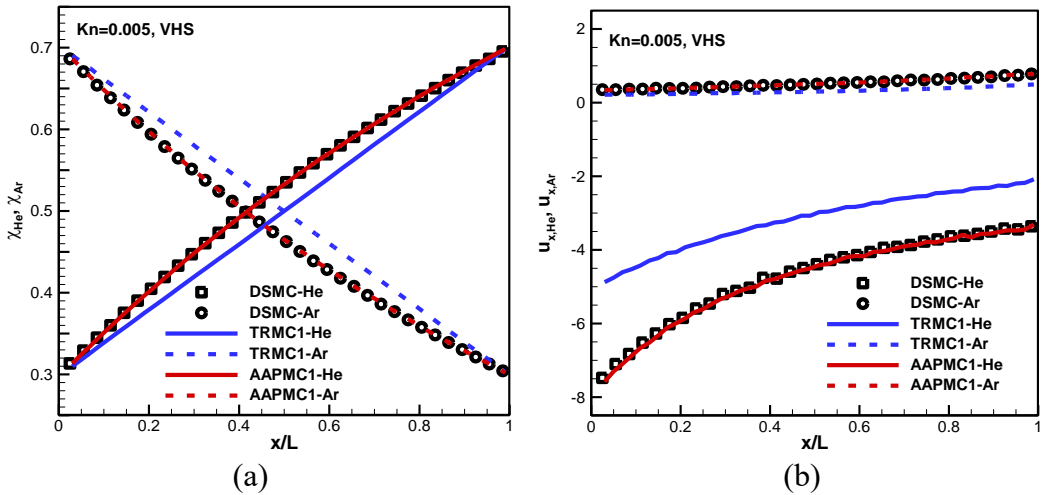

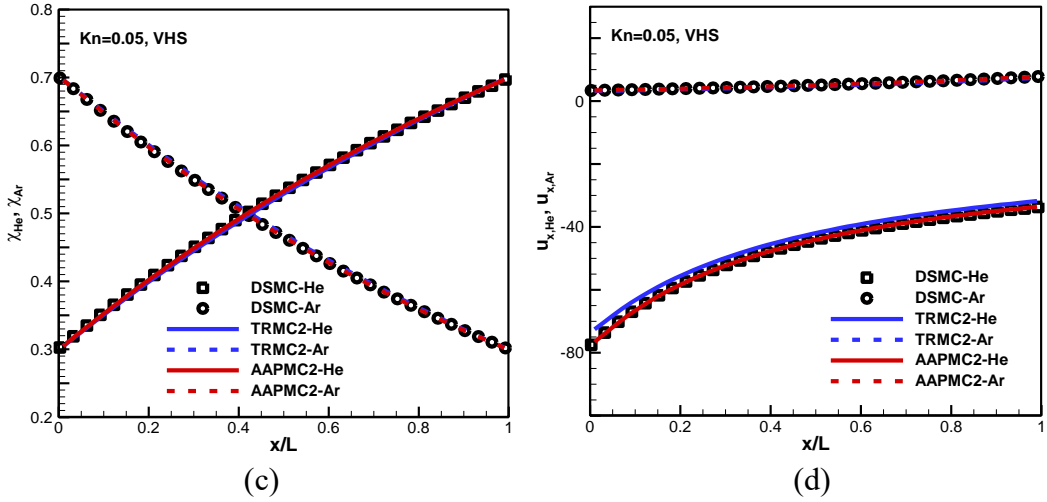

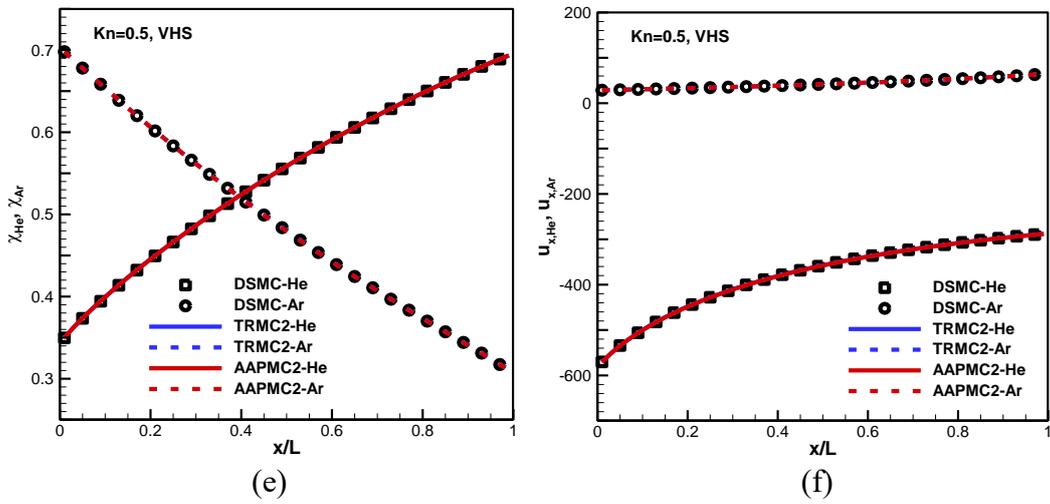

Figure 4. Comparison of concentration gradients and diffusion velocities in He-Ar mixture with VHS model at Kn numbers of 0.5, 0.05 and 0.005.

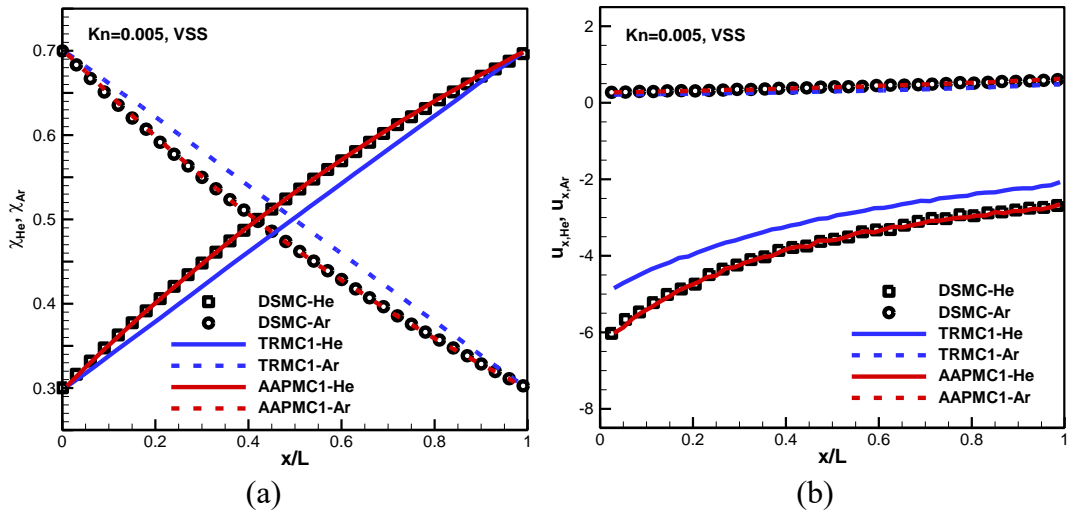

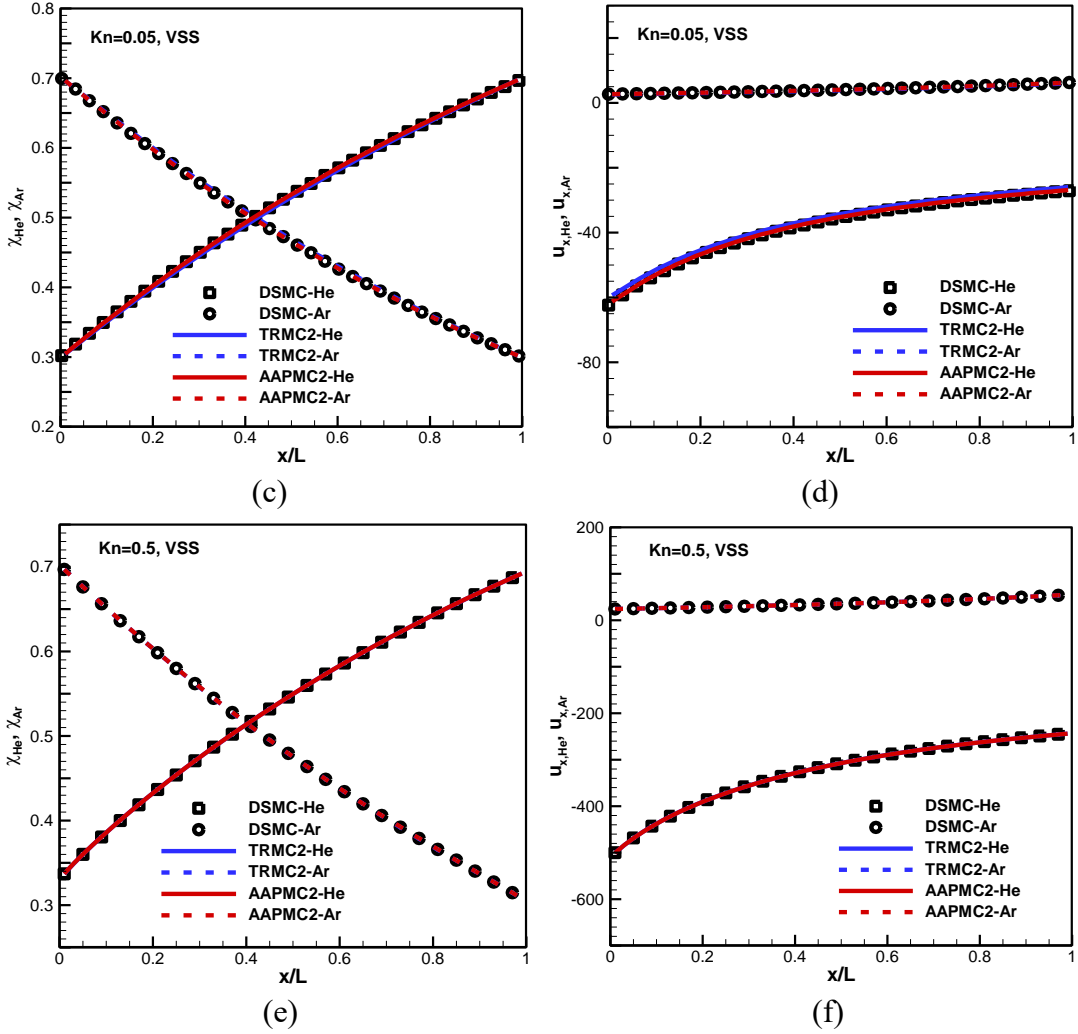

Figure 5. Comparison of concentration gradients and diffusion velocities in He-Ar mixture with VSS model at Kn numbers of 0.5, 0.05 and 0.005.

**5.2 Fourier flow**

The Fourier flows are generated by a temperature difference between two parallel plates. In our simulations, the temperatures of the upper and lower plates keep $323K$ and $223\ K$, respectively. Two mixtures are calculated, one is the binary mixture of helium-argon and the other one is the ternary mixture of helium-argon-xenon. For both mixtures, the initial temperature and total number density are set to $273K$ and $2.689 \times 10^{25} m^{-3}$. The initial species concentrations are specified as $\chi_{He} = 0.5$ and $\chi_{Ar} = 0.5$ for the helium-argon mixture, and $\chi_{He} = 0.3$, $\chi_{Ar} = 0.4$ and $\chi_{Xe} = 0.3$ for the helium-argon-xenon mixture. Under these initial conditions, the mean free paths for the binary and ternary mixtures are calculated to be $8.30 \times 10^{-8}\ m$ and $5.85 \times 10^{-8}\ m$, respectively. Three Knudsen numbers from the rarefied to continuum flow regime are considered, i.e., Kn=0.5, 0.05 and 0.005. The computational domain is defined as $[0, H]$, and fully diffusive boundary conditions are employed. The DSMC, TRMC and AAPMC methods are implemented using the VHS model. In DSMC, 50, 200 and 2000 uniform grids are utilized, while the TRMC and AAPMC methods only used 50 uniform grids for all cases. Similar to the calculation of mass diffusion, the CFL number is set to 0.3 for all simulations.

Figure 6 demonstrates the species concentration and temperature profiles for the He-Ar mixture. For Kn=0.005, the cell size and time step of the TRMC and AAPMC methods are nearly 4 times the initial mean free path and 1.2 times the initial mean collision time, respectively. Under this situation, the traditional TRMC method fails to capture the thermal diffusion phenomenon, as depicted in Fig. 6(a). In contrast, the AAPMC method yields more precise results due to the AP property of the NS equations. As the Kn number increases, both the TRMC and AAPMC methods approach the results obtained by DSMC. In Figs. 6(c)- 6(f), these three methods provide consistent species concentration and temperature profiles. In Figure 7, for all Kn numbers, the heat fluxes of the AAPMC method are in close agreement with those of DSMC. However, due to the AP property of the Euler equations, the TRMC method obtains lower heat fluxes in the continuum flow regime. In addition, the second-order TRMC and AAPMC methods seem more accurate than the first-order ones in the rarefied flow regime, as shown in Figs. 7(b) and 7(c). Moreover, in the continuum flow regime, the difference between the first and second-order schemes becomes negligible due to the use of the same macro collision operator.

Figure 8 illustrates the species concentration and temperature profiles for the He-Ar-Xe mixture. The AAPMC method also agrees well with DSMC across all Kn numbers. Conversely, the traditional TRMC method is less accurate in the continuum flow regime. In addition, Fig. 9 depicts the heat fluxes for the ternary mixture. Similar to the findings from the binary mixture, the AAPMC method yields superior results compared to the TRMC method at small Kn numbers. The numerical simulation for the Fourier flows demonstrates that the AAPMC method can be easily extended to gas mixtures with more than two species like the DSMC method.

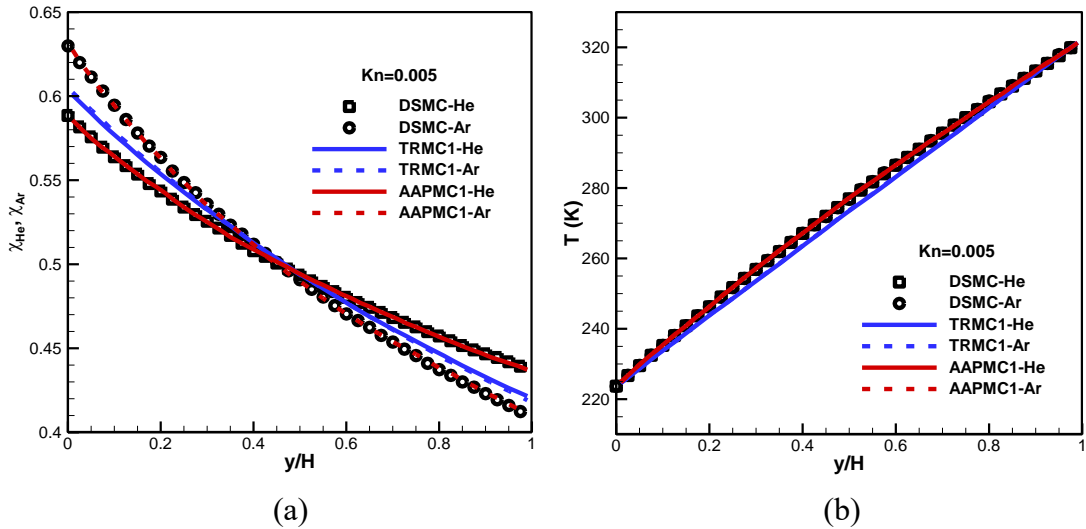

(a)            (b)

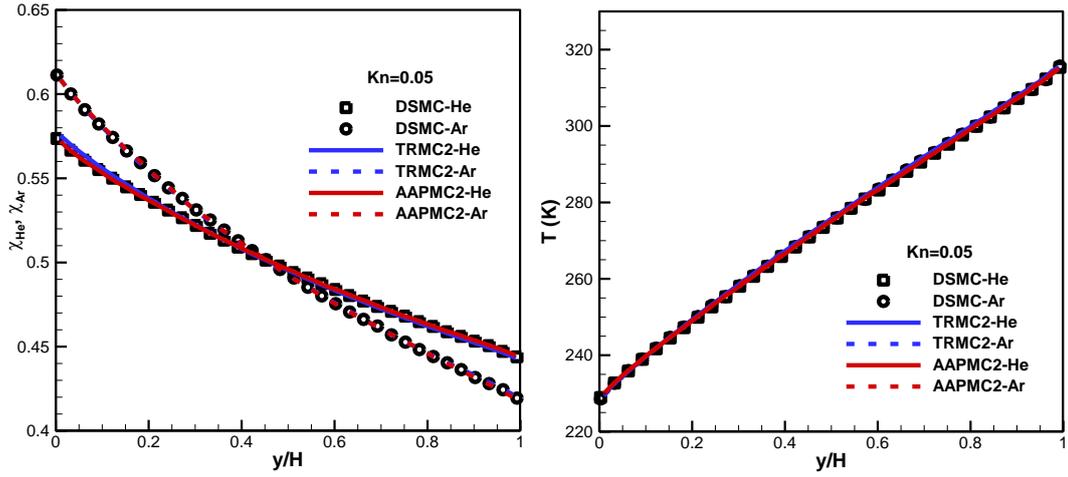

(c)  (d)

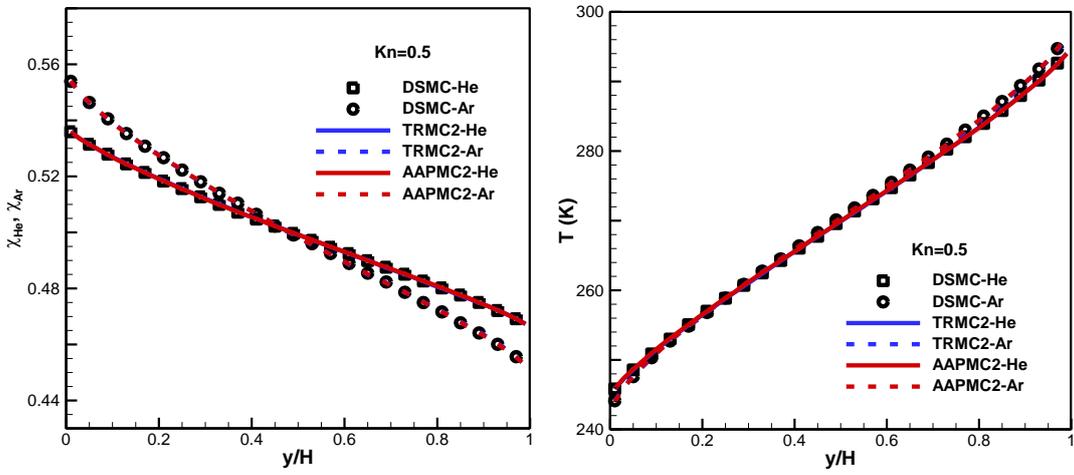

(e)  (f)

Figure 6. Comparison of concentration gradients and temperatures in Fourier flows of a He-Ar mixture at Kn numbers of 0.5, 0.05 and 0.005.

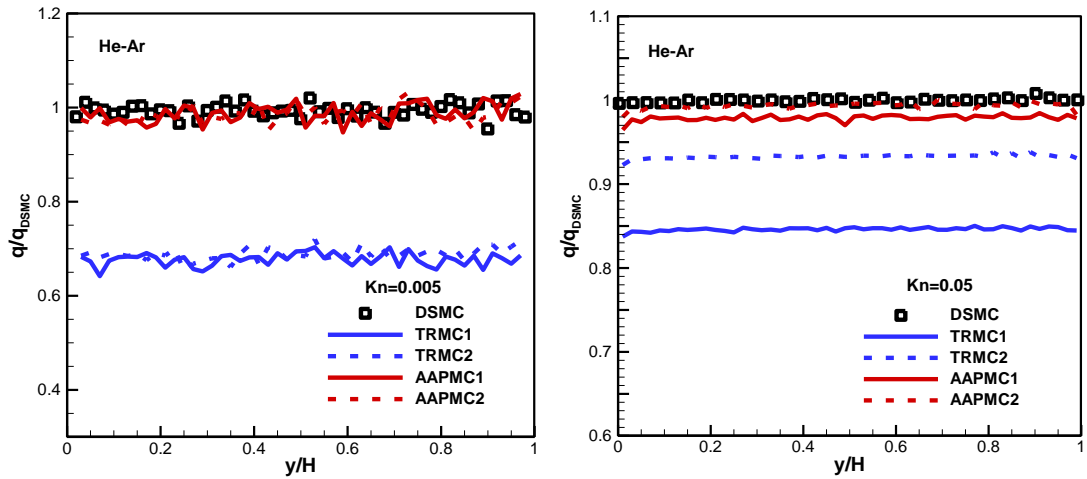

(a)  (b)

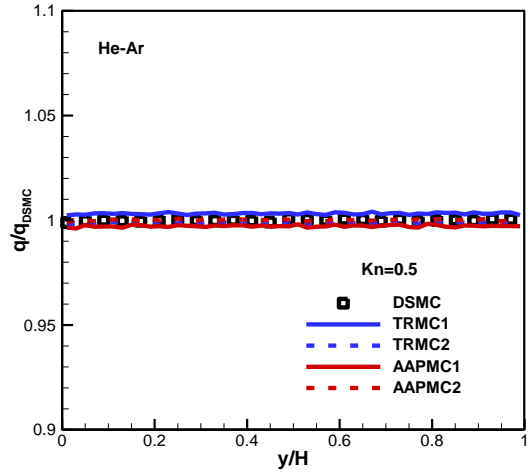

(c)

Figure 7. Comparison of heat fluxes in Fourier flows of a He-Ar mixture at Kn numbers of 0.5, 0.05 and 0.005.

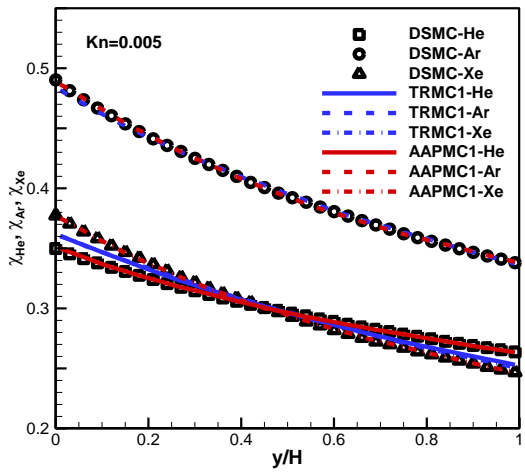 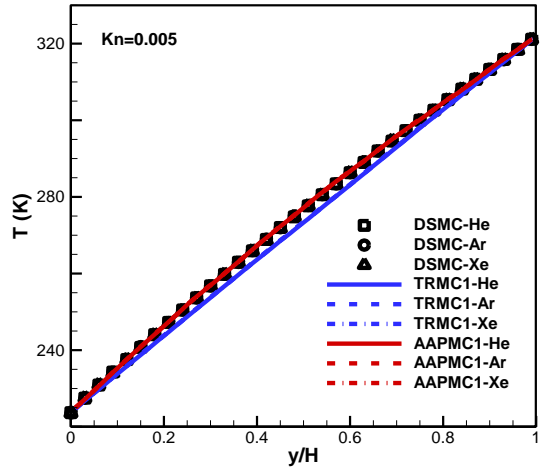

(a) (b)

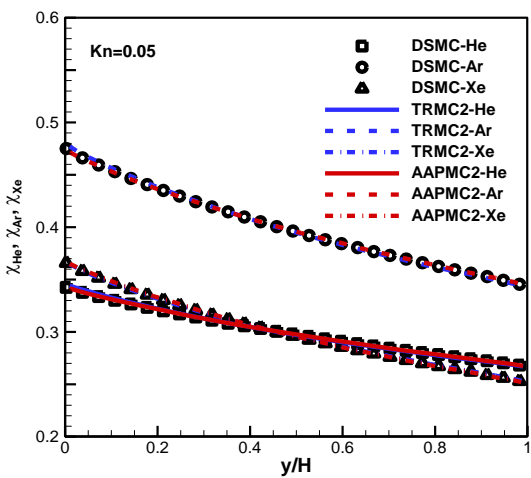 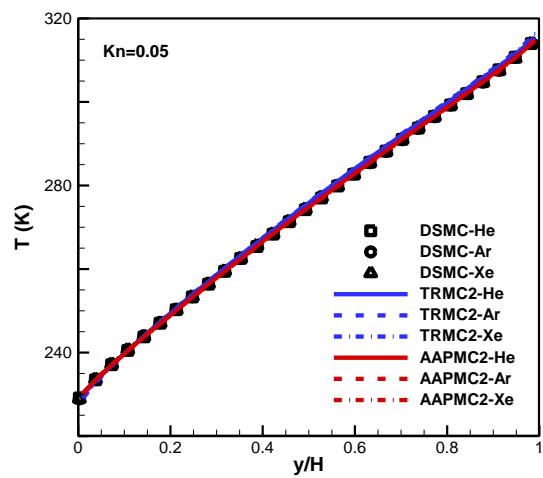

(c) (d)

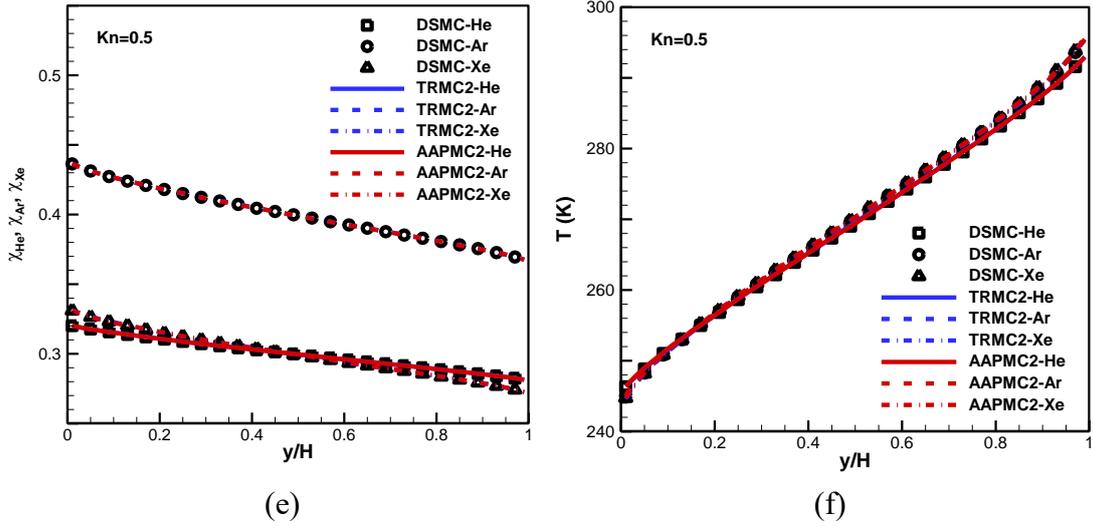

(e)                            (f)

Figure 8. Comparison of concentration gradients and temperatures in Fourier flows of a He-Ar-Xe mixture at Kn numbers of 0.5, 0.05 and 0.005.

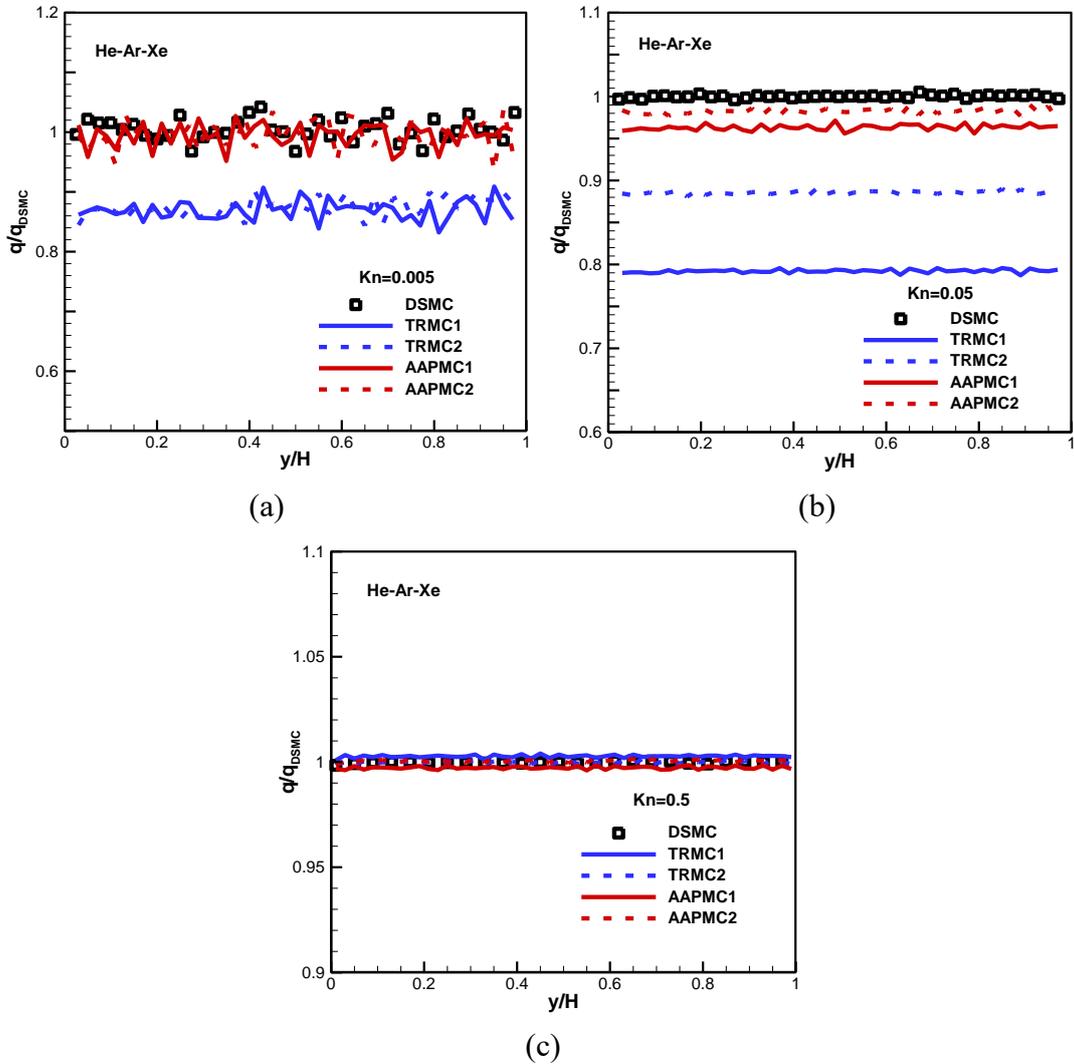

Figure 9. Comparison of heat fluxes in Fourier flows of a He-Ar-Xe mixture at Kn numbers of 0.5, 0.05 and 0.005.

**5.3 Couette flow**

The Couette flows are confined between two parallel plates that move in opposite directions. The upper and lower plates have velocities of $300\ m/s$ and $-300\ m/s$, respectively. Both plates are maintained at a constant temperature of $273\ K$, where fully diffusive boundary conditions are applied. Initially, the mixture of helium-argon-xenon is at standard condition ($T_0 = 273K$ and $p_0 = 1\ atm$), with species concentrations set to $\chi_{He} = 0.3$, $\chi_{Ar} = 0.4$ and $\chi_{Xe} = 0.3$. Three Knudsen numbers, specifically 0.5, 0.05 and 0.005, are calculated using the DSMC, TRMC and AAPMC methods with the VHS model. Other computational parameters are consistent with those used in the simulations of Fourier flows.

Figure 10 presents comparisons of species velocity and temperatures. The AAPMC method obtains consistent results with DSMC in all Kn numbers. However, the TRMC method underestimates the viscous heating and temperature at small Kn numbers with the same grids, as shown in Fig. 10(b) and 10(d). For Kn numbers of 0.05 and 0.5, the He-Ar-Xe mixture exhibits a temperature difference attributable to the dispersed masses of these species. The nonequilibrium parameters for the mixture are illustrated in Figure 11(a). We note that the nonequilibrium parameters $B_{Xe}$ and $B_{Ar}$ are greater than $B_{He}$, indicating a higher degree of nonequilibrium for the heavier species. Moreover, Figure 10(d) depicts that the temperature of the light species He aligns with DSMC results, however, temperature profiles for Ar and Xe from AAPMC2 are lower than the reference data. In this case, since $B_{Xe} > B_{Ar} > 0.1$, the DSMC collision operator is performed for Xe and Ar in the coupling scheme. As DSMC demonstrates a lower accuracy than the AAPMC collision operator, refiner grids are required to achieve the same level of resolution. Therefore, we observe that these temperature predictions can be improved by employing a refined grid with 100 cells, as shown in Fig 11(b).

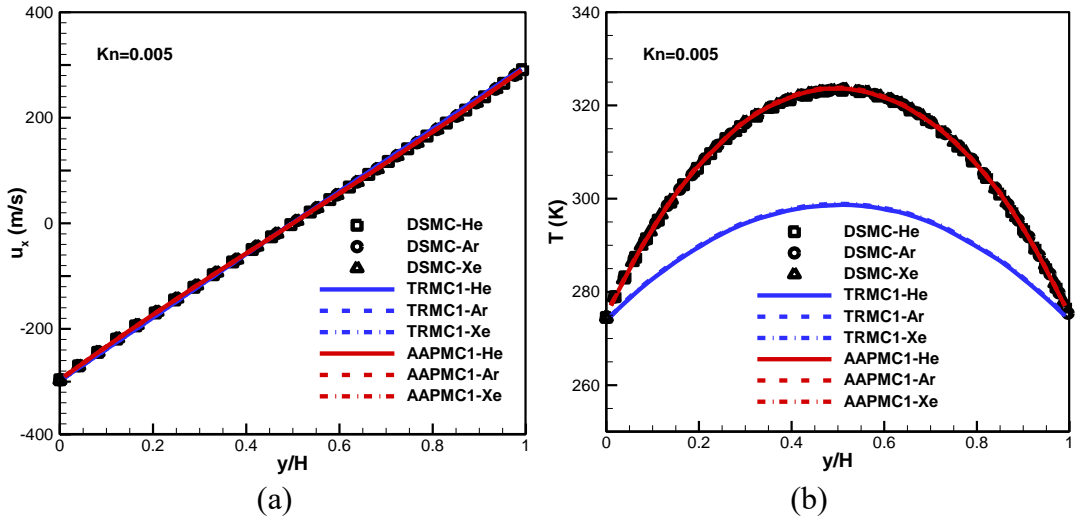

(a)　　　　　　　　　　　　　　(b)

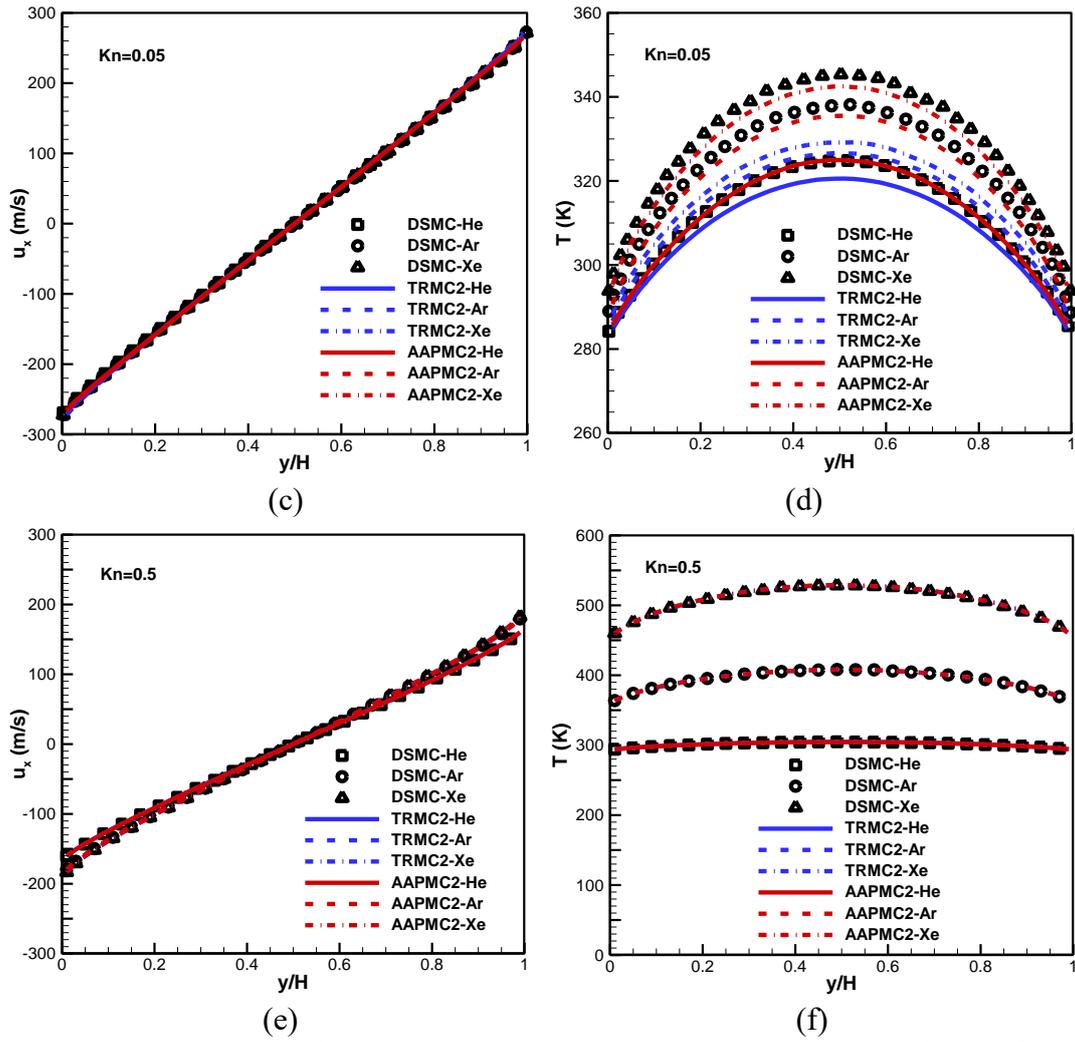

Figure 10. Comparison of species velocity and temperatures in Couette flows of a He-Ar-Xe mixture at Kn numbers of 0.5, 0.05 and 0.005.

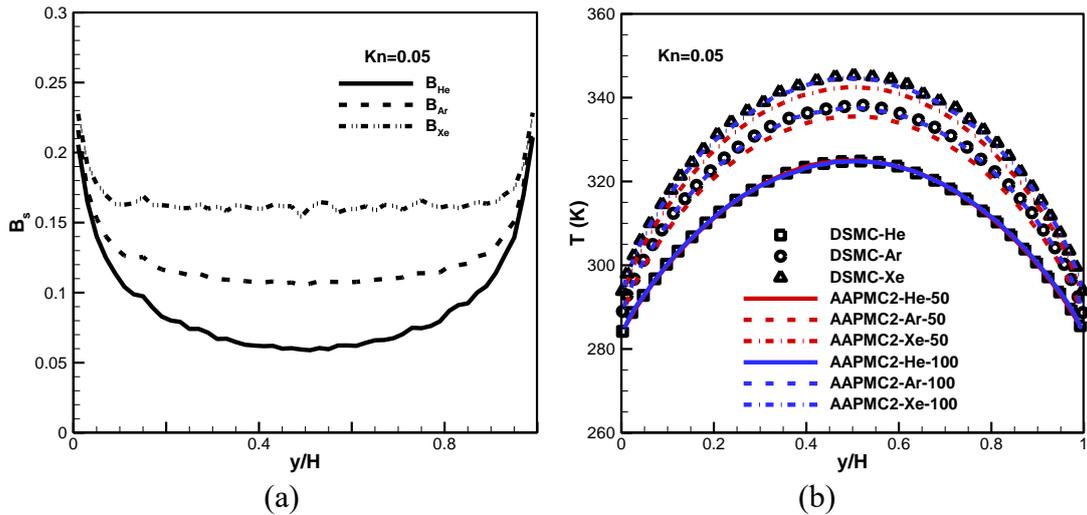

Figure 11. (a) The nonequilibrium parameters and (b) species temperatures for refined grids for the Couette flow at Kn =0.05 using the DAPMC method.

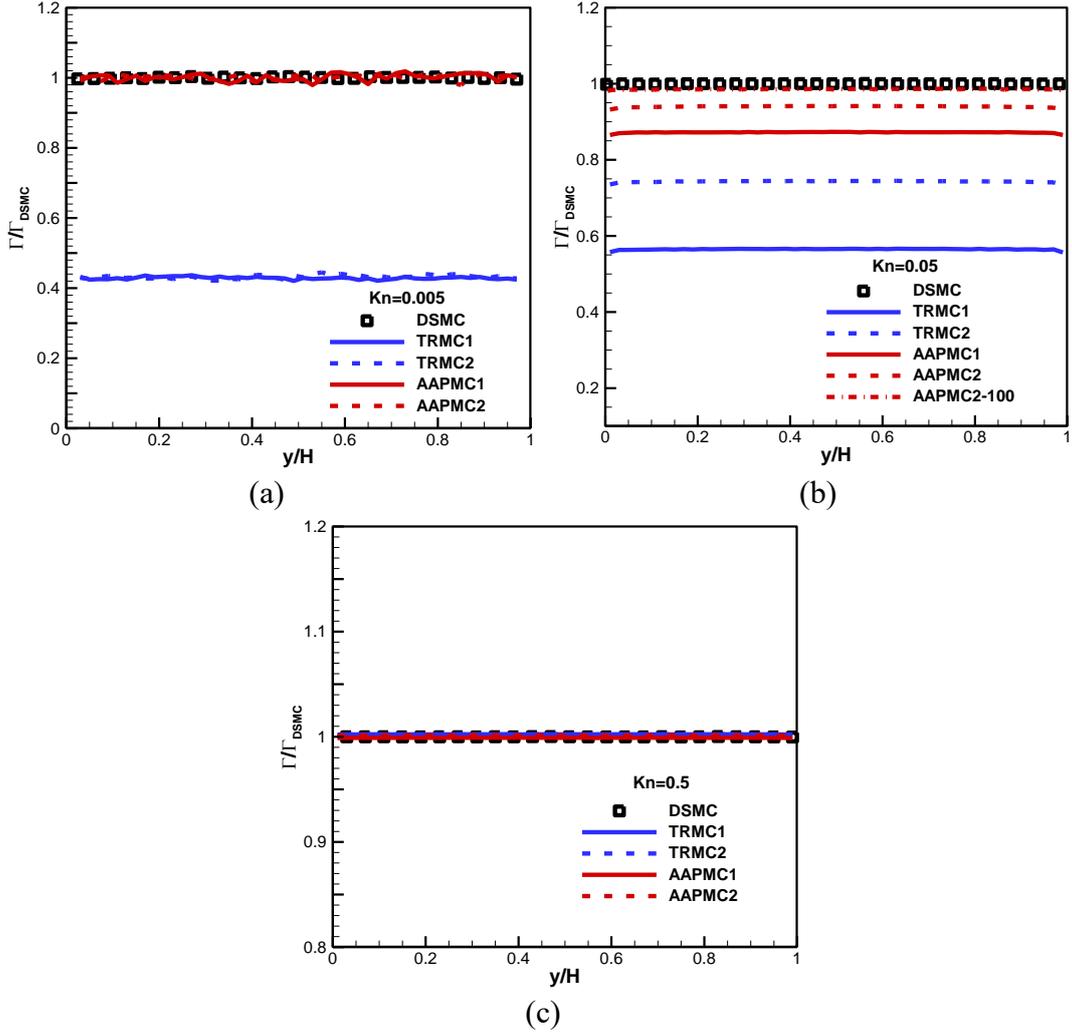

Figure 12. Comparison of shear stress in Couette flows of a He-Ar-Xe mixture at Kn numbers of 0.5, 0.05 and 0.005.

Figure 12 compares the shear stress for the three cases. Similar to the observations in Fourier flows, we note that the AAPMC method is more accurate than the TRMC method, and the second-order scheme performs more effectively in the rarefied flow regime. Additionally, figure 12(b) presents the shear stress calculated using the AAPMC2 scheme with 100 cells. This provides a better prediction than that obtained with half the number of cells, similar to the temperature results in Fig. 11(b).

**5.4 Shock structure**

Two types of mixture are considered in the shock wave flows. The first is a mixture of He and Ar with species concentrations set to $\chi_{He} = \chi_{Ar} = 0.5$; the second is a mixture of He, Ar and Xe with species concentrations set to $\chi_{He} = 0.9$, $\chi_{Ar} = 0.05$ and $\chi_{Xe} = 0.05$ in the free stream. In both cases, the free stream's temperature and total number density are $300K$ and $1.6095 \times 10^{21} m^{-3}$, respectively, and the Mach number is 3.0. The computational domain spans $L = 100\lambda_\infty$, where $\lambda_\infty$ denotes the mean free path of the free stream. We compare the results of the DSMC, TRMC, and AAPMC methods with the VHS model. For the DSMC method, 500 uniform cells are employed, while one-tenth of cells are used in the TRMC and AAPMC methods. The CFL number is set to 0.5.

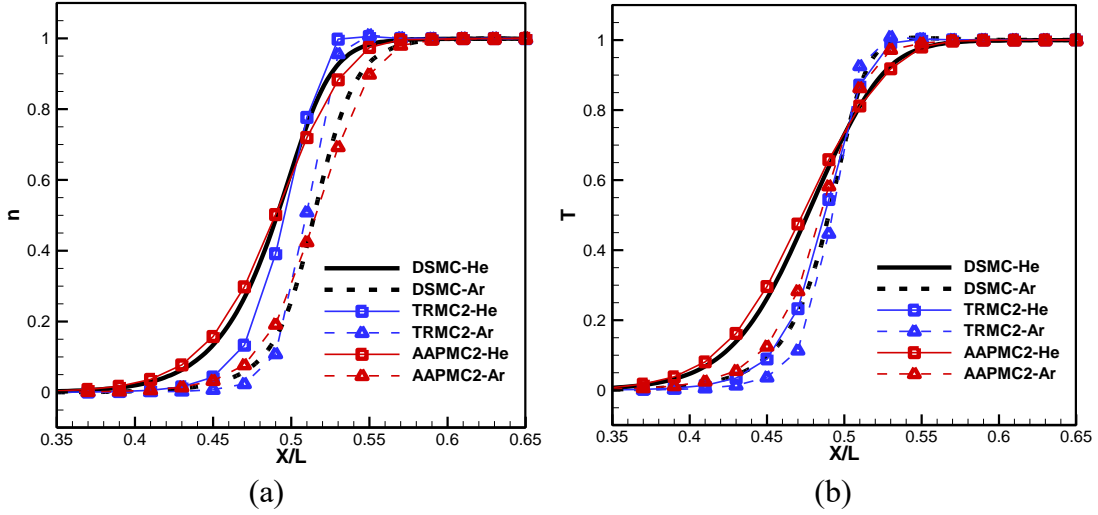

(a)  (b)

Figure 13. Comparison of the species number density and temperature profiles of the shock wave at Ma =3 for the binary gas mixture of He and Ar.

Figures 13 and 14 present the normalized species number density and temperature distributions relative to their values ahead of and behind the shock wave. For both the binary and ternary gas mixtures, the AAPMC method agrees with DSMC. However, due to the AP property of the Euler equations, the TRMC method yields a thinner shock structure. Given that $m_{Xe}/m_{He} \approx 30$, the significant temperature difference is observed in the He-Ar-Xe mixture, which can be accurately reproduced by the AAPMC method even with fewer grids. Moreover, Figure 15 shows the species velocities and nonequilibrium parameters for the ternary gas mixture of He-Ar-Xe. It is observed that the thickness of the shock wave and the nonequilibrium parameter $B_s$ are larger for the heavy species, indicating a higher degree of non-equilibrium. Additionally, $B_s > 0.1$ in the shock area for all species. Therefore, the AAPMC-DSMC coupling scheme employs the DSMC collision solver in the shock area and the standard AAPMC collision solver ahead of and behind the shock. This coupling approach makes the AAPMC method preferable for accurately and efficiently calculating multi-scale gas flows.

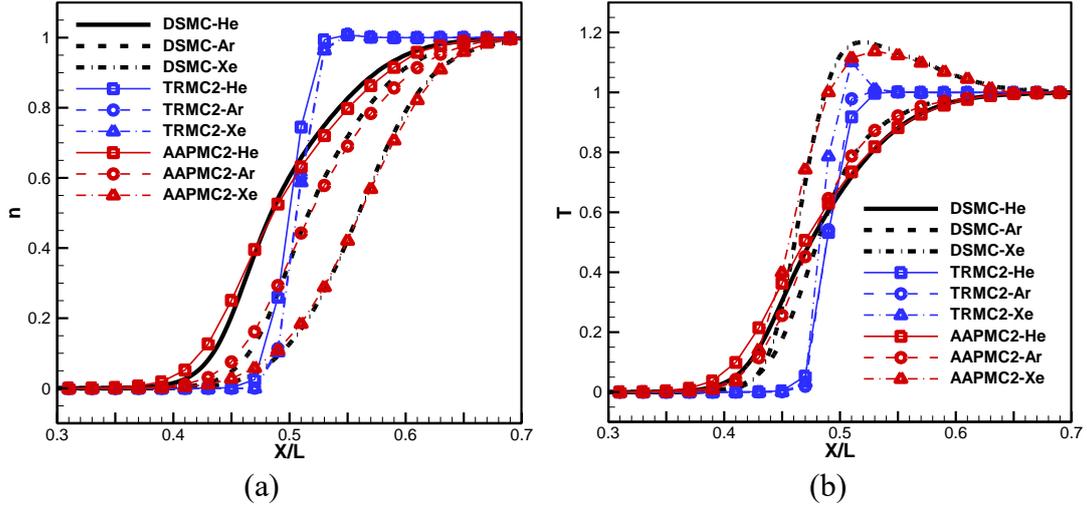

Figure 14. Comparison of the species number density and temperature profiles of the shock wave at Ma =3 for the ternary gas mixture of He, Ar and Xe.

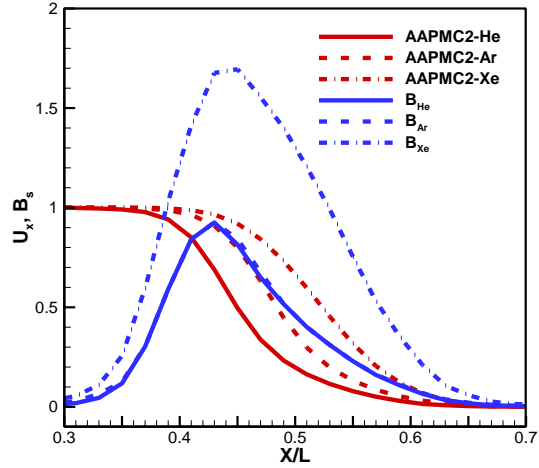

Figure 15. The species velocities and nonequilibrium parameters of the shock wave at Ma =3 for the ternary gas mixture of He, Ar and Xe.

**5.5 Poiseuille flow**

The Poiseuille flows are calculated using the hard sphere model for a He-Ar gas mixture. The initial number densities of He and Ar are equal to $0.5n_0$. Three Knudsen numbers, i.e., 0.2, 0.02 and 0.002, are considered. Consequently, the distance between the upper and lower plates are $5\lambda_0$, $50\lambda_0$ and $500\lambda_0$, respectively, where $\lambda_0$ denotes the mean free path of the mixture under the initial condition of $T_0 = 273K$ and $p_0 = 1atm$, thus $\lambda_0 = 10^{-7}m$. The corresponding pressure gradients are chosen as $6 \times 10^{10}$, $6 \times 10^8$ and $6 \times 10^6 \, Pa \, m^{-1}$, respectively, for three Kn numbers. During simulations, the upper and lower plates are kept static with a constant temperature of $273K$.

Figure 16 compares the velocity distributions of the TRMC, AAPMC and DSMC methods for the three Knudsen numbers. The DSMC method utilizes 40, 400 and 4000 uniform grids for Kn=0.2, 0.02 and 0.002, respectively. In contrast, fewer grids are employed in the TRMC and AAPMC methods, i.e., 40 grids for Kn=0.2, 80 grids for Kn=0.02 and 100 grids for Kn=0.002. The results in Figure 16 demonstrate that the AAPMC method is more accurate. Furthermore, Figure 17 depicts the $L_2$-norm of error for the mean velocity at Kn=0.002, confirming that the AAPMC method can achieve second-order accuracy in the fluid limit. Here, the reference solution $u_x^{ref}$ is derived from the NS solution.

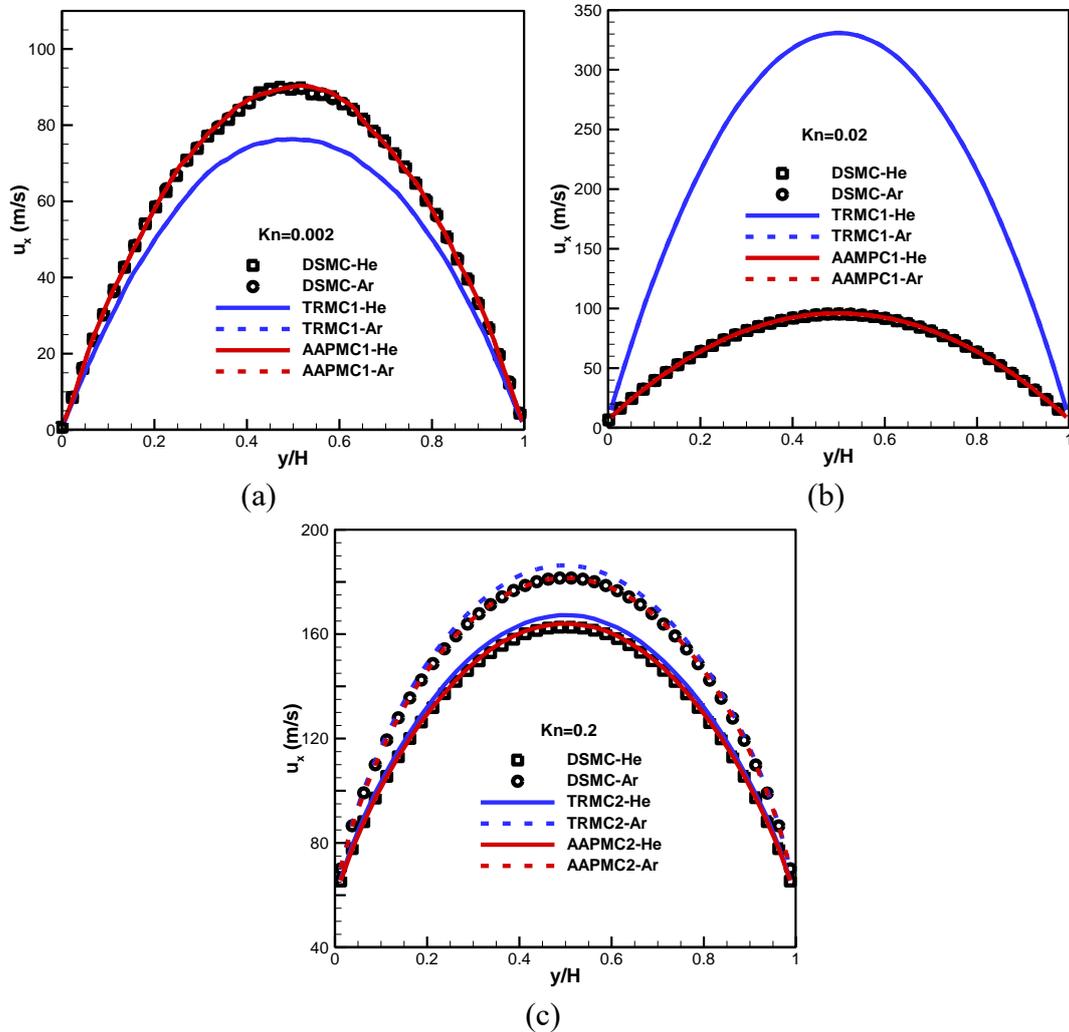

Figure 16. Comparison of the velocity distributions of the Poiseuille flows at Kn numbers of 0.2, 0.02 and 0.002.

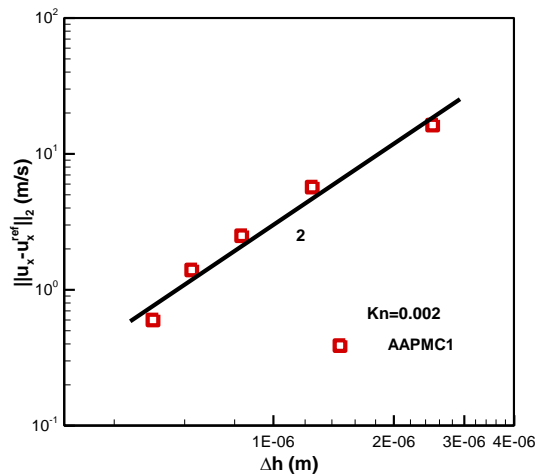

Figure 17. $L_2$-norm of error for the mean velocity of the AAPMC method at Kn=0.002.

## 6. Conclusion

In this paper, a new asymptotic-preserving Monte Carlo method has been developed for simulating multi-species gas flows. By employing Strang splitting and micro-macro decomposition techniques on the collision operator of the traditional APMC methods, this proposed scheme not only satisfies the asymptotic-preserving property of the NS equations but also achieves second-order accuracy in the fluid limit. These enhancements significantly improve the resolution for the transport processes and flow structures in the continuum flow regime, making this proposed APMC method particularly advantageous for simulating multi-scale gas flows. Furthermore, by utilizing a local DSMC collision solver, we directly compute the first-order CE expansions for multi-species gases, avoiding complex modeling for interactions among multi-species. Consequently, this approach is well-suited for extension to more complex gas flows, such as chemical reaction flows. Additionally, coupled with DSMC, this new APMC scheme can be extended for the simulations of multi-species gases with disparate masses. Moreover, although this work is based on the TRMC method, extending this strategy to other APMC schemes, such as the ExpRK and AP-DSMC methods, is feasible and straightforward.

## Appendix A The AAPMC method based on the second-order TRMC scheme

The collision operator of the second-order TRMC method is given by

$$f_s^{n+1}(V;x) = \mathcal{C}_{\Delta t}^{TRMC2}(f_s^*) = A_0 f_s^* + A_1 \frac{P_s(f^*,f^*)}{\beta} + A_2 \frac{P_s(f^*,f^1)}{\beta} + A_3 M_{s,eq}^*, \quad (A1)$$

where $A_0 = e^{-\beta \Delta t/\varepsilon}$, $A_1 = e^{-\beta \Delta t/\varepsilon}(1 - e^{-\beta \Delta t/\varepsilon})$, $A_2 = 1 - \sum_{i=0}^{1} A_i - A_3$ and $A_3 = (1 - e^{-\beta \Delta t/\varepsilon})^4$, and $f_s^1 = \frac{1}{\beta} P_s(f^*, f^*)$. Employing Eq. (4.3), we can also construct the AAPMC method based on the second-order TRMC scheme. The macro collision part remains unchanged; however, using Eq. (A1), the micro collision part is written as

$$\mathcal{C}_{\Delta t}^{TRMC2}(f_s^*) - \mathcal{C}_{\Delta t}^{TRMC2,(1)}(f_s^*) = \begin{cases} A_0 f_s^* + A_1 \frac{P_s(f^*,f^*)}{\beta} + A_2 \frac{P_s(f^*,f^1)}{\beta} + A_3 M_{s,eq}^* \\ -\left[ \begin{array}{c} (1 - A_3) f_s^{n+1,(1)} \\ -\left(\frac{\Delta t}{2\varepsilon} - \frac{1}{\beta} A_1 - \frac{3}{2\beta} A_2 - \frac{\Delta t}{2\varepsilon} A_3\right) Q_s^{(1)}(f^{n+1}) \\ +\left(\left(\frac{1}{2\beta} - \frac{3\Delta t}{4\varepsilon}\right) \frac{1}{\beta} A_2 - \frac{\Delta t}{2\varepsilon} \frac{1}{\beta} A_1\right) Q_s^{(1)}\left[Q_s^{(1)}(f^{n+1})\right] \end{array} \right] \end{cases}. \quad (A2)$$

where the higher order term, i.e., $Q_s^{(1)}\{Q_s^{(1)}[Q_s^{(1)}(f^{n+1})]\}$, is neglected.

Combining Eqs. (A2), (4.3) and (4.9), the collision operator of the AAPMC method based on the second-order TRMC method is obtained as

$$\mathcal{C}_{\Delta t}^{AAPMC2}(f_s^*) = A_0 f_s^* + A_1 \frac{P_s(f^*,f^*)}{\beta} + A_2 \frac{P_s(f^*,f^1)}{\beta} + A_3(M_{s,eq}^{n+1} + f_s^\alpha), \quad (A3)$$

with

$$f_s^\alpha = \begin{cases} f_s^{n+1,(1)} + \left(\frac{\Delta t}{\varepsilon} - \frac{1}{\beta}A_1 - \frac{3}{2\beta}A_2 - \frac{\Delta t}{2\varepsilon}A_3\right)\frac{1}{A_3}Q_s^{(1)}(f^{n+1}) \\ + \left(\frac{\Delta t}{2\varepsilon}\frac{1}{\beta}A_1 - \left(\frac{1}{2\beta} - \frac{3\Delta t}{4\varepsilon}\right)\frac{1}{\beta}A_2\right)\frac{1}{A_3}Q_s^{(1)}\left[Q_s^{(1)}(f^{n+1})\right] \end{cases}. \tag{A4}$$

The sampling algorithm for $f_s^\alpha$ is identical to that used for the first-order method. Moreover, the implementation of the third term on the right-hand side of Eq. (A3) is the same as that of the second-order TRMC methods and can be found in reference [44].

## Appendix B The deviational DSMC algorithm for the linear Boltzmann Equation

Generally, the PDF of velocities can be decomposed into a Maxwellian and a deviational distribution as

$$f_s = M_s + d_s = M_s + d_s^p - d_s^n, \tag{B1}$$

where $d_s$ denotes the deviational distribution. In the standard DSMC method, simulated particles are employed to represent the PDF $f_s$, and all the particle weights are set to be positive. In contrast, the deviational DSMC algorithm simulates only the deviational distribution. Therefore, particle weights in this algorithm can be either positive or negative, and their respective PDFs are denoted as $d_s^p$ and $d_s^n$. $d_s^p$ and $d_s^n$ are defined as $N_{dev}^{(P)} = \frac{\mathbb{V}}{W_p}\int d_s^p(V)dV$ and $N_{dev}^{(N)} = \frac{\mathbb{V}}{W_p}\int d_s^n(V)dV$. In addition, since the Maxwellian distribution is deterministic, statistic noise can be significantly reduced. As a result, the deviational DSMC algorithm was initially developed for the low variance DSMC method [47], and has since been applied to various variance reduction problems [48-50].

Substituting Eq. (B1) into the Boltzmann equation (2.1), we can rewrite the homogenous linear Boltzmann equation as

$$\frac{\partial f_s}{\partial t} = \frac{1}{\varepsilon}\sum_{k=1}^{N_s}[Q^+(M_k, d_s) + Q^+(d_k, M_s) - Q^-(M_k)d_s - Q^-(d_k)M_s], \tag{B2}$$

where $Q_s(M, M) = 0$ and the nonlinear terms have been omitted.

Furthermore, by splitting $f_s$ into three parts as shown in Eq. (B1), we decompose Eq. (B2) into three equations, each accounting for $M_s$, $d_s^p$ and $d_s^n$ respectively [49], i.e.,

$$\begin{cases} \frac{\partial M_s}{\partial t} = 0 \\ \frac{\partial d_s^p}{\partial t} = \frac{1}{\varepsilon}\sum_{k=1}^{N_s}[Q^+(M_k, d_s^p) + Q^+(d_k^p, M_s) + Q^-(d_k^n)M_s - Q^-(M_k)d_s^p] \\ \frac{\partial d_s^n}{\partial t} = \frac{1}{\varepsilon}\sum_{k=1}^{N_s}[Q^+(M_k, d_s^n) + Q^+(d_k^n, M_s) + Q^-(d_k^p)M_s - Q^-(M_k)d_s^n] \end{cases}. \tag{B3}$$

The first equation indicates that the Maxwellian distribution remains invariant. The second and third equations depict the evolution of the PDF for the positive and negative particles, respectively. Moreover, the first three terms in these two equations correspond to the gain terms, while the last term in each corresponds to the loss term.

According to Eq. (B3), particle collisions can be categorized into P-M and N-M collisions. The former refers to collisions between positive and Maxwellian particles, while the latter involves collisions between negative and Maxwellian particles. B. Yan, et. al. [49] have provided detailed collision rules for the P-M and N-M collisions. This deviational DSMC algorithm can be directly applied to multi-species gas flows. Here, we only explain the P-M collision process; the N-M collision process can be derived similarly.

**P-M collision.** Collecting all terms involving P-M collisions in Eq. (B3), we have

$$\begin{cases} \frac{\partial M_s}{\partial t}\Big|_{PM} = 0 \\ \frac{\partial d_s^p}{\partial t}\Big|_{PM} = \frac{1}{\varepsilon}\sum_{k=1}^{N_s}[Q^+(M_k, d_s^p) + Q^+(d_k^p, M_s) - Q^-(M_k)d_s^p] \\ \frac{\partial d_s^n}{\partial t}\Big|_{PM} = \frac{1}{\varepsilon}\sum_{k=1}^{N_s}[Q^-(d_k^p)M_s] \end{cases} \quad (B4)$$

This collision process can be described as

$$M_s, V_{s,+} \to M_s, V'_{s,+}, W'_{s,+}, W_{s,-}. \quad (B5)$$

Equation (B5) implies that a positive particle with velocity $V_{s,+}$ of species $s$ first undergoes a collision with a particle of velocity $W_k$, which is sampled from the Maxwellian distribution of species $k$. Secondly, their post-collisional velocities, denoted by $V'_s$ and $W'_k$, are determined based on the dynamics of binary collisions. Thirdly, the original positive particle with velocity $V_{s,+}$ is eliminated, while a new negative particle with velocity $W_{k,-}$ and two new positive particles with velocity $V'_{s,+}$ and $W'_{k,+}$ are created. Similarly, the N-M collision process is described by

$$M_s, V_{s,-} \to M_s, V'_{s,-}, W'_{s,-}, W_{s,+}. \quad (B6)$$

## Acknowledgments

This work was supported by the Interdisciplinary Research Program of Hust (No. 2024JCYJ003).## Reference